\documentclass[11pt]{article}
\usepackage[utf8]{inputenc}
\usepackage[T1]{fontenc}
\usepackage{lmodern,microtype}
\usepackage[letterpaper,margin=1in]{geometry}
\usepackage{amsmath,amssymb,amsthm,mathtools,bm}
\usepackage{thmtools}
\usepackage{graphicx,booktabs,array}
\usepackage[inline]{enumitem}
\usepackage{algorithm}
\usepackage[noEnd=true,indLines=false]{algpseudocodex}
\usepackage{xcolor}
\usepackage{colortbl}
\usepackage{hyperref}
\usepackage[nameinlink,noabbrev,capitalize]{cleveref}
\AddToHook{env/algorithmic/begin}{\expandafter\def\csname @currentcounter\endcsname{ALG@line}}
\crefalias{ALG@line}{line}
\makeatletter
\providecommand*{\theHALG@line}{}
\renewcommand*{\theHALG@line}{\arabic{algorithm}.\arabic{ALG@line}}
\makeatother
\definecolor{linkblue}{RGB}{0,102,204}
\definecolor{citegreen}{RGB}{0,140,70}
\definecolor{urlorange}{RGB}{230,110,0}
\hypersetup{
	colorlinks=true,
	linktoc=page,
	bookmarksnumbered=true,
	bookmarksopen=true,
	bookmarksopenlevel=1,
	bookmarksdepth=2,
	pdfdisplaydoctitle=true,
	linkcolor=linkblue,
	citecolor=citegreen,
	urlcolor=urlorange
}
\numberwithin{equation}{section}

\setlist[itemize,enumerate]{topsep=4pt,itemsep=2pt,parsep=0pt}
\algrenewcommand\algorithmicrequire{\textbf{Input:}}
\algrenewcommand\algorithmicensure{\textbf{Output:}}

\makeatletter
\@ifundefined{newcounteralias}{}{%
	\renewcommand\thmt@autorefsetup{\@xa\def\csname\thmt@envname autorefname\@xa\endcsname\@xa{\thmt@thmname}}%
}
\makeatother

\declaretheorem[name=Theorem,numberwithin=section,style=plain]{theorem}
\declaretheorem[name=Lemma,sibling=theorem,style=plain]{lemma}

\declaretheorem[name=Claim,numbered=no,style=plain]{claim*}

\declaretheorem[name=Corollary,sibling=theorem,style=plain]{corollary}
\declaretheorem[name=Definition,sibling=theorem,style=definition]{definition}

\declaretheorem[name=Remark,sibling=theorem,style=remark]{remark}

\newcommand{\vb}{\bm{b}}
\newcommand{\one}{\mathbf{1}}
\newcommand{\OPT}{\mathrm{OPT}}
\DeclareMathOperator{\rk}{rk}
\DeclareMathOperator{\cl}{cl}
\newcommand{\defi}{\operatorname{def}}

\usepackage[colorinlistoftodos,textsize=footnotesize]{todonotes}

\newcommand{\degin}{\deg^{\text{in}}}

\providecommand{\dotcup}{\mathbin{\dot\cup}}
\providecommand{\head}{\operatorname{head}}
\providecommand{\cT}{\mathcal{T}}
\providecommand{\Ix}{I^{\mathrm{cr}}}

\usepackage{tikz}
\usetikzlibrary{arrows.meta,decorations.pathmorphing,decorations.pathreplacing,calc,positioning,shapes.geometric}
\providecommand{\FG}{D_{\mathrm{fine}}}
\providecommand{\CG}{D_{\mathrm{cmp}}}

\title{Minimum-Cost $k$-Arborescence Packing in $\tilde O(km+k^4 n)$ Time}
\author{Zhaienhe Zhou\thanks{University of Michigan, United States. Email: \texttt{zhaienhezhou@gmail.com}.
 Work done in part while the author was an undergraduate student at the University of Science and Technology of China.
  }
}
\date{}
\hypersetup{
    pdftitle={Minimum-Cost k-Arborescence Packing in O\textasciitilde(km+k\textasciicircum4 n) Time},
    pdfauthor={Zhaienhe Zhou},
    pdflang={en-US}
}

\begin{document}
\maketitle

\pagenumbering{gobble}

\begin{abstract}
  We give a deterministic, strongly polynomial algorithm running in $\tilde{O}(km+k^4n)$ time that finds $k$ arc-disjoint arborescences of minimum total cost in a directed multigraph with $n$ vertices and $m$ arcs. This significantly improves upon the state of the art for $k \ge 2$ with $k=n^{o(1)}$: the prior fastest weakly polynomial and strongly polynomial algorithms take $\tilde{O}(km\sqrt{n}\log C)$  and $\tilde{O}(kmn)$ time (Gabow and Xu, 1989; Gabow, 1995), respectively.

	Roughly, our technique can be viewed as a way to apply a linear number of \emph{shortest} augmenting paths in near-linear total time for the special case of the weighted matroid intersection problem between the $k$-fold graphic matroid and the indegree partition matroid. This was previously done in the unweighted case by Gabow (1995), but our approach is different.

	As an extension, we also obtain a deterministic $\tilde{O}(km^{1+o(1)}+k^5n)$-time algorithm for minimum-cost maximum-cardinality packing of $k$ forests in undirected graphs. In particular, this finds $k$ edge-disjoint spanning trees of minimum total cost, if they exist. The previous fastest algorithm takes $\tilde{O}(kn^2+m)$ time (Gabow and Westermann, 1992).
\end{abstract}

\newpage

\tableofcontents

\newpage
\pagenumbering{arabic}

\section{Introduction}\label{sec:intro}

One of the central goals of network design is to guarantee connectivity at low cost despite arc failures. In a directed graph, a packing of $k$ arc-disjoint spanning arborescences rooted at a common vertex guarantees that every vertex remains reachable from the root after the deletion of any $k-1$ arcs.

We therefore study the problem of \emph{minimum-cost $k$-arborescence packing}. Formally, let $G=(V,E)$ be a directed multigraph with $n=|V|$ and $m=|E|$, let $r\in V$ be the root, let $k\ge1$ be an integer, and let $c:E\to\mathbb R$ be arc costs. An \emph{$r$-arborescence} is a spanning tree of $G$ whose arcs are directed away from $r$. Our goal is to find $k$ arc-disjoint $r$-arborescences in $G$ of minimum total cost. In this work, we mainly consider the small-$k$ regime.

For $k=1$, the classical minimum-cost arborescence problem~\cite{chuShortestArborescence1965,edmondsOptimumBranchings1967,tarjanFindingOptimumBranchings1977,gabowEfficientAlgorithmsFinding1986} can be solved in $O(m+n\log n)$ time.
For fixed $k\ge2$, near-linear $\tilde{O}(km)$-time\footnote{We use $\widetilde O$ to hide factors polylogarithmic in $n,m,k$.} algorithms are known for the unweighted problem~\cite{gabowMatroidApproachFinding1995,arkhipovFasterAlgorithmsPacking2025}, whereas all previously known weighted algorithms incur an additional polynomial factor in $n$.

Namely, for integer costs in $[1,C]$, Gabow and Xu's algorithm%
\footnote{Gabow stated in his paper~\cite{gabowMatroidApproachFinding1995} that, using the techniques in~\cite{gabowIndependentAssignments1989}, one can obtain running times of both $\tilde{O}(knm)$ and $\tilde{O}(k\sqrt n m \log C)$ for $k$-arborescence packing, with the latter based on the cost-scaling framework of~\cite{gabowEfficientTheoreticPractical1996}.%
}%
~\cite{gabowIndependentAssignments1989,gabowMatroidApproachFinding1995} runs in $\widetilde O(k\sqrt n\,m\log C)$ time by combining a modification of the matroid intersection algorithm of Frank~\cite{frankWeightedMatroidIntersection1981} with a cost-scaling framework. For general real costs, the best known algorithm still needs $\tilde{O}(knm)$ time~\cite{gabowIndependentAssignments1989,gabowMatroidApproachFinding1995}.

Minimum-cost $k$-arborescence packing is a key building block in approximation algorithms for network design, including edge-connected spanning subgraphs~\cite{khullerBiconnectivityApproximationsGraph1994,chalermsookApproximatingKEdgeConnected2022} and flexible graph connectivity~\cite{boydApproximationAlgorithmsFlexible2021,chandrasekaranApproximationFlexibleGraph2026}, and it is sometimes the bottleneck of their running time \cite{khullerBiconnectivityApproximationsGraph1994,chalermsookApproximatingKEdgeConnected2022}. This raises the question of whether faster or even near-linear time algorithms exist.

In this work, we answer this affirmatively by giving a deterministic, strongly polynomial algorithm for arbitrary real costs that runs in near-linear time for any polylogarithmic $k$:

\begin{theorem}\label{thm:main}
	Let $n\ge2$. A minimum-cost packing of $k$ arc-disjoint $r$-arborescences, or a report that none exists, can be computed deterministically in
	\begin{equation}\label{eq:main-time}
		O\bigl(km\log(kn)+k^2n\log n\log^2(kn)+k^4n\log^2(kn)\bigr)
	\end{equation}
	time for arbitrary real costs. For fixed $k$, this is $O(m\log n+n\log^3n)$.
\end{theorem}

The main difficulty is to batch many augmentations while preserving optimality. The earlier integer-cost algorithm~\cite{gabowIndependentAssignments1989,gabowEfficientTheoreticPractical1996} already batches augmentations within its cost-scaling framework, but the number of batch rounds is $O (\sqrt{n})$, incurring an extra $\sqrt{n}$ factor in the running time.

We develop a batching framework that exploits the locality
of shortest augmenting paths to execute many more augmentations in each round, so that the total number of batch rounds is $O(k\log(kn))$.
This approach was already successful in the unweighted case \cite{gabowMatroidApproachFinding1995}, but there are many obstacles to applying it in the weighted case, as described in \cref{sec:overview}.

\paragraph{Minimum-cost $k$-forest packing.}

We further extend our algorithm to the \emph{minimum-cost $k$-forest packing} problem, the undirected analogue of $k$-arborescence packing. Here we compute a minimum-cost maximum-cardinality union of $k$ edge-disjoint forests in an undirected multigraph.
This problem subsumes, for example, finding $k$ edge-disjoint spanning trees of minimum total cost, if they exist.

For $k=1$, this is essentially the minimum spanning tree problem (assuming $G$ is connected), with numerous classical deterministic near-linear algorithms~\cite{boruvkaJistemProblemuMinimalnim1926,kruskalShortestSpanningSubtree1956,primShortestConnectionNetworks1957,gabowEfficientAlgorithmsFinding1986}. For $k\ge2$, the unweighted variant can be solved in $\widetilde O(m+k^{3/2}n\sqrt m)$ time~\cite{gabowForestsFramesGames1992} and in $\widetilde O(m+(kn)^{3/2})$ time~\cite{blikstadFastAlgorithmsDynamicOracle2023,quanrudFasterPackingCovering2024}. More recently, Arkhipov and Kolmogorov~\cite{arkhipovFasterAlgorithmsPacking2025} obtained a $\widetilde O(k^3\min\{kn,m\}+km^{1+o(1)})$-time algorithm, achieving almost-linear time for $k=n^{o(1)}$.

For arbitrary real costs, minimum-cost packing of $k$ edge-disjoint spanning trees (instead of forests) can be solved in $O(m\log m+k^2n^2)$ time~\cite{roskindMinimumCostSpanningTrees1985}. Later, Gabow and Westermann gave an algorithm for the general $k$-forest problem running in $\widetilde O(kn^2+m)$ time~\cite{gabowForestsFramesGames1992}.

Our algorithm achieves almost-linear time for arbitrary real costs and $k=n^{o(1)}$.

\begin{restatable}[Undirected forest packing]{theorem}{undirectedforestpacking}
	\label{thm:undirected-forest-packing}
	Let $G$ be an undirected multigraph with $n\ge2$ vertices, $m$ edges, and arbitrary real edge costs, and let $k\ge1$ be an integer. A minimum-cost maximum-cardinality union of $k$ edge-disjoint forests, together with a forest decomposition, can be computed deterministically in
	\begin{equation}\label{eq:undirected-modular-time}
		O\Bigl(k\log(kn)\cdot\mathsf{MCF}\bigl(O(m),O(m)\bigr)
		+k^3n\log n\log^3(kn)+k^5n\log^3(kn)\Bigr)
	\end{equation}
	time.
\end{restatable}

Here $\mathsf{MCF}(N,M)$ denotes the running time of a deterministic algorithm that finds an integral minimum-cost maximum flow in a network with $N$ vertices and $M$ arcs, whose costs and capacities are integers bounded by a polynomial in $N+M$. Plugging in the algorithm of~\cite{vandenbrandDeterministicAlmostLinear2023} gives a strongly polynomial algorithm running in $\tilde{O}(km^{1+o(1)}+k^5n)$ time, which is almost linear for $k=n^{o(1)}$.

\begin{table}[htbp]
	\centering\small
	\begin{tabular}{@{}lll@{}}
		\toprule
		Costs              & Running time                                & Reference                                                                       \\
		\midrule
		\multicolumn{3}{@{}l}{\emph{$k$-Arborescence packing}}                                                                                             \\[2pt]
		real, $k=1$        & $\widetilde O(m)$                           & \cite{tarjanFindingOptimumBranchings1977,gabowEfficientAlgorithmsFinding1986}   \\
		unweighted         & $\widetilde O(km)$                          & \cite{gabowMatroidApproachFinding1995}                                          \\
		real               & $\widetilde O(knm)$                         & \cite{gabowIndependentAssignments1989,gabowMatroidApproachFinding1995}          \\
		integer in $[1,C]$ & $\widetilde O(k\sqrt n\,m\log C)$           & \cite{gabowIndependentAssignments1989,gabowMatroidApproachFinding1995}          \\
		\rowcolor{linkblue!8}
		real               & $\widetilde O(km+k^4n)$                     & \textbf{\Cref{thm:main}}                                                        \\[3pt]
		\midrule
		\multicolumn{3}{@{}l}{\emph{Undirected $k$-forest packing}}                                                                                        \\[2pt]
		unweighted         & $\widetilde O(m+k^{3/2}n\sqrt{m})$          & \cite{gabowForestsFramesGames1992}                                              \\
		unweighted         & $\widetilde O(m+(kn)^{3/2})$                & \cite{blikstadFastAlgorithmsDynamicOracle2023,quanrudFasterPackingCovering2024} \\
		unweighted         & $\widetilde O(k^3\min\{kn,m\}+km^{1+o(1)})$ & \cite{arkhipovFasterAlgorithmsPacking2025}                                      \\
		real               & $\widetilde O(kn^2+m)$                      & \cite{gabowForestsFramesGames1992}                                              \\
		\rowcolor{linkblue!8}
		real               & $\widetilde O(km^{1+o(1)}+k^5n)$            & \textbf{\Cref{thm:undirected-forest-packing}}                                   \\
		\bottomrule
	\end{tabular}
	\caption{Algorithms for $k$-arborescence packing and $k$-forest packing.}
	\label{tab:running-times}
\end{table}

\paragraph{Applications.} We give two further applications of \Cref{thm:main} in \Cref{sec:applications}.

\emph{Network design.} For nonnegative costs, we obtain deterministic $2$-approximation algorithms (\Cref{cor:network-design}) with the running time stated in~\eqref{eq:main-time} for the minimum-cost $k$-edge-connected spanning subgraph of an undirected graph and the minimum-cost strongly $k$-edge-connected spanning subgraph of a digraph. The undirected case uses the reduction of Khuller and Vishkin~\cite{khullerBiconnectivityApproximationsGraph1994}. For the undirected problem, a previous randomized $(2+\varepsilon)$-approximation algorithm runs in $\widetilde O((m+k^2n^{3/2})/\varepsilon^2)$ time~\cite{chalermsookApproximatingKEdgeConnected2022}.

Another application is an algorithm for a generalization of the minimum-cost $k$-arborescence packing problem, called \emph{minimum-cost indegree-bounded forest packing}, with the same running time as in~\eqref{eq:main-time} (\Cref{thm:indegree-bounded-forest-packing}).

\paragraph{Organization.} \Cref{sec:overview} gives an overview of our techniques, and \Cref{sec:prelim} collects preliminaries. \Cref{sec:arborescence-packing} develops the batching framework, the compressed search graph, and the compatible-set computation. Using black-box negative-weight shortest-path algorithms, we obtain a weaker version of \Cref{thm:main}: for fixed $k$, near-linear time with logarithmic cost dependence for integer costs, and strongly polynomial almost-linear time for real costs (\Cref{rem:black-box-shortest-paths}). In \Cref{sec:vertex-prices}, we construct a feasible potential function and use Dijkstra's algorithm to prove \Cref{thm:main}. \Cref{sec:applications} gives the applications, including minimum-cost undirected forest packing and network design.

\section{Overview}\label{sec:overview}

By Edmonds' theorem (\Cref{lem:arborescence-formulation}), an arc set forms a $k$-arborescence packing if and only if it can be partitioned into $k$ forests of the underlying undirected graph and has indegree $k$ at every vertex other than $r$. The first condition is equivalent to independence in the union $\mathcal{G}^k$ of $k$ graphic matroids, and the second corresponds to independence in a partition matroid whose classes consist of arcs sharing the same head. Therefore, our problem can be formulated as a weighted matroid intersection problem. Throughout this section, \emph{independent} refers specifically to independence in $\mathcal{G}^k$.

The standard algorithm for weighted matroid intersection~\cite{lawlerMatroidIntersectionAlgorithms1975} searches for augmenting paths in an exchange graph, whose vertices represent ground set elements and whose directed edges represent exchanges allowed by the two matroids. Along such a path, an augmentation alternately inserts and removes elements, increasing the size of the common independent set by one. In our setting, the solution has size $k(n-1)$, requiring the same number of augmentations. Rebuilding the exchange graph and recomputing shortest paths after each augmentation yields at least a quadratic running time in the size of the solution.

The main idea of this paper is a way to perform a linear number of augmentations in near-linear total time for this weighted matroid intersection problem. The only earlier method we know that does this is Gabow's algorithm for the unweighted problem~\cite{gabowMatroidApproachFinding1995}. We first recall why his algorithm is fast, and then explain the obstacles that costs create and how we overcome them.

\paragraph{Gabow's algorithm: augmenting in local time.}

At a high level, Gabow's algorithm maintains a set $I$ that is independent in $\mathcal{G}^i$, along with a forest decomposition $I = F_1 \dotcup \dots \dotcup F_i$, where $F_1, \dots, F_{i-1}$ are spanning trees and $F_i$ is a forest. The set $I$ grows until $F_i$ becomes a spanning tree, and then the algorithm proceeds to pack the next tree.

His algorithm uses what we call a \emph{fine-grained augmenting path}, which is more fine-grained than standard matroid-intersection augmenting paths in the exchange graph: each exchange in $\mathcal{G}^i$ is expanded into a sequence of swaps among the maintained forests, allowing the forest decomposition to be updated explicitly. Gabow shows that these paths have useful locality: each path stays entirely within a single component of $F_i$, except for its final arc.

In each round, every component of $F_i$ has such a path starting from a vertex in it with indegree less than $i$. By locality, finding each of them (with cyclic-scanning search) only requires time linear in the size of the containing component, so the total time still sums to linear. To apply many paths at once, Gabow applies a simple cycle-avoiding selection rule to choose a subset of them, keeping the resulting $I$ independent. He shows that each round reduces the number of components of $F_i$ by half, so after $O(\log n)$ rounds, $F_i$ becomes a spanning tree. This shares the same spirit as Bor\r{u}vka's algorithm~\cite{boruvkaJistemProblemuMinimalnim1926} for MST.

\paragraph{Keeping the augmentations cheapest (\Cref{sec:relax-and-augment}).}

Gabow's way of packing the trees one by one can be seen as relaxing the matroid: once $I$ is a complete solution for $\mathcal G^i$, the algorithm switches from $\mathcal G^i$ to $\mathcal G^{i+1}$. Without costs this is harmless. With costs, the augmenting-path algorithm needs $I$ to be cheapest among the common independent sets of its size (\Cref{thm:weighted-matroid-intersection}). This can fail: a cheapest packing of $i$ trees need not be cheapest for $\mathcal G^{i+1}$, and we do not know how to repair $I$ efficiently.

We therefore fix $\mathcal{G}^k$ from the start, so that all $k$ forests grow together, and raise only the indegree bounds. Our new invariant is that $I$ is cheapest among all independent sets with the same indegree vector as $I$. Raising an indegree bound does not change the sets $I$  is compared with, so the invariant still holds. Switching from $\mathcal G^i$ to $\mathcal G^{i+1}$, in contrast, adds new sets to compare with, even for the same indegree vector. The relaxation theorem below shows that augmentations also keep the invariant.

Concretely, let $\OPT(\vb)$ be the minimum cost of an independent set with indegree vector $\vb$. We maintain the invariant that $I$ has cost $\OPT(\vb)$, where $\vb$ is the indegree vector of $I$. We also maintain a forest decomposition $I=F_1\dotcup\cdots\dotcup F_k$. %
To raise the indegree of a vertex $u$, we use a \emph{matroid-level augmenting path from $u$}: a standard augmenting path in the exchange graph of $\mathcal G^k$ and the current indegree bounds, but starting with an arc entering $u$. It ends with a \emph{free} arc $e$, that is, $e\notin I$ and $I+e$ is independent. We call it matroid-level to distinguish it from Gabow's fine-grained paths.

Our first contribution is an elementary proof of a relaxation theorem for an arbitrary matroid, here $\mathcal{G}^k$, intersected with a partition matroid (\Cref{thm:relaxation}). Let $\Delta_u=\OPT(\vb+\one_u)-\OPT(\vb)$. First, a shortest augmenting path from $u$ (with ties broken by the number of exchanges) raises the indegree of $u$ at cost $\Delta_u$, so the augmented set costs exactly $\OPT(\vb+\one_u)$, maintaining the invariant. Second, raising the indegrees of all vertices in a set $U$ together costs at least the sum of the single-vertex costs:
\[
	\OPT(\vb+\one_U)\ge \OPT(\vb)+\sum_{u\in U}\Delta_u.
\]
So if element-disjoint shortest paths, one for each $u\in U$, can be executed together, the result costs the right-hand side and is thus optimal (\Cref{lem:batch-optimality}). This does not need a path to remain shortest after another path is executed, as in usual augmenting-path arguments.

\paragraph{Locality to tight blocks (\Cref{sec:localized-proposals}).}

Gabow's fine-grained paths are local to the components of $F_i$, guaranteeing local-time search and path disjointness. Since we fix $\mathcal G^k$ and grow all $k$ forests together, there is no single forest whose components play this role.

We find that the right notion is locality to tight blocks. A vertex set $T$ is \emph{tight} if $I$ has $k(|T|-1)$ arcs inside $T$, the most that $k$ forests allow. The maximal tight sets, the \emph{tight blocks}, partition $V$.
A shortest (matroid-level) augmenting path from a vertex $u$ stays inside the tight block of $u$, except for its last arc, which enters the block from outside (\Cref{lem:proposal-localization}).
This is because such a path stops at the first free arc, and every arc outside $I$ crossing tight blocks is free (\Cref{lem:free-elements}). This locality is more refined than Gabow's $F_i$-component locality, since tight blocks refine the components of every forest. It also depends only on $I$ instead of the decomposition.

\paragraph{Selecting compatible augmenting paths (\Cref{sec:contracted-compatibility}).}
Call a vertex $v\ne r$ \emph{deficient} if its indegree is less than $k$, and a tight block deficient if it contains one. Each phase of our algorithm picks one deficient vertex $u_T$ in every deficient tight block $T$, and computes a shortest augmenting path from $u_T$, the \emph{proposal} of $T$. By locality, the proposal changes only arcs inside $T$, except for its last arc, the \emph{terminal}, which enters $T$, and proposals of different blocks are element-disjoint.

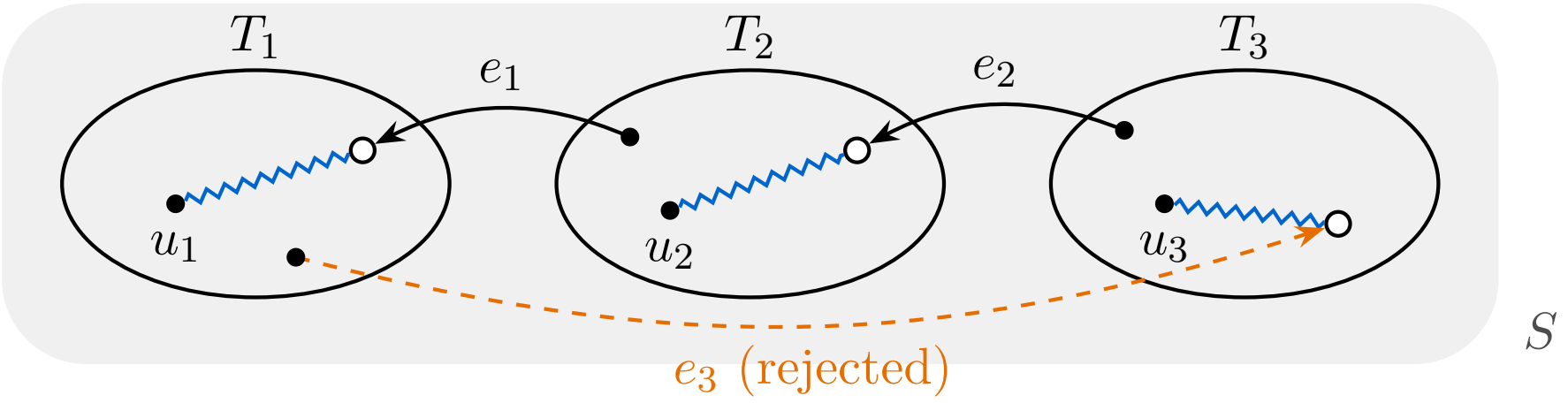
\begin{figure}[t]
	\centering
	\begin{tikzpicture}[
			vtx/.style={circle,fill=black,inner sep=1.4pt},
			hv/.style={circle,draw=black,thick,fill=white,inner sep=1.8pt},
			blk/.style={draw,thick,ellipse,minimum width=2.9cm,minimum height=1.7cm},
			term/.style={-{Stealth[length=2.2mm]},thick},
			zig/.style={decorate,decoration={zigzag,segment length=4pt,amplitude=1.2pt},thick,linkblue}]
		\fill[gray!12,rounded corners=18pt] (-1.9,-1.35) rectangle (9.3,1.35);
		\node[blk] (T1) at (0,0) {};
		\node[blk] (T2) at (3.7,0) {};
		\node[blk] (T3) at (7.4,0) {};
		\node at (0,1.1) {$T_1$}; \node at (3.7,1.1) {$T_2$}; \node at (7.4,1.1) {$T_3$};
		\node[vtx,label=below:{$u_1$}] (u1) at (-0.6,-0.15) {};
		\node[hv] (h1) at (0.8,0.25) {};
		\draw[zig] (u1) -- (h1);
		\node[vtx,label=below:{$u_2$}] (u2) at (3.1,-0.2) {};
		\node[hv] (h2) at (4.5,0.25) {};
		\draw[zig] (u2) -- (h2);
		\node[vtx,label=below:{$u_3$}] (u3) at (6.8,-0.15) {};
		\node[hv] (h3) at (8.1,-0.3) {};
		\draw[zig] (u3) -- (h3);
		\coordinate (t1) at (2.8,0.35);
		\coordinate (t2) at (6.5,0.4);
		\coordinate (t3) at (0.3,-0.55);
		\draw[term] (t1) node[vtx]{} to[bend right=25] node[above,pos=0.5]{$e_1$} (h1);
		\draw[term] (t2) node[vtx]{} to[bend right=25] node[above,pos=0.5]{$e_2$} (h2);
		\draw[term,dashed,urlorange] (t3) node[vtx]{} to[bend right=16] node[below,pos=0.5]{$e_3$ (rejected)} (h3);
		\node[gray!60!black,anchor=west] at (9.35,-1.1) {$S$};
	\end{tikzpicture}
	\caption{One phase for $k=1$. Each proposal changes the tree inside its tight block $T_j$ (blue zigzag) and adds a terminal $e_j$ entering $T_j$. We execute the proposals of $T_1$ and $T_2$ and reject $e_3$ (orange dashed arc): after $e_1,e_2$ are added, the shaded set $S$ is tight and contains both ends of $e_3$.}
	\label{fig:batching-phase}
\end{figure}

Still, executing all of them may break independence. We next determine which paths can be executed together.
Intuitively, the change inside $T$ keeps the arcs inside $T$ independent and $T$ tight, and it does not affect anything outside $T$. So we contract every tight block, and let $\Ix$ be the arcs of $I$ that cross the blocks. We show that a set of proposals can be executed together if and only if $\Ix$ together with their terminals is a union of $k$ forests in the contracted graph (\Cref{lem:contraction-lifting}), regardless of how they change the inside of the blocks. We call such a set of terminals \emph{compatible}. So we choose a maximal compatible set of terminals $A'$, and then execute the corresponding proposals at once (\Cref{fig:batching-phase}). By the relaxation theorem, the result again satisfies the invariant.

For $k=1$, the tight blocks are the components of the forest $I$ and $\Ix=\varnothing$, so a compatible set is a set of terminals with no cycle in the contracted graph, as in Gabow's cycle-avoiding rule. For $k\ge2$ this rule no longer works: the terminals must fit into $k$ forests together with $\Ix$. Equivalently, letting $A$ be the set of all terminals, we must extend $\Ix$ to a basis of $\Ix\cup A$ in $\mathcal G^k$ of the contracted graph. We call the unweighted forest-packing algorithm of Arkhipov and Kolmogorov~\cite{arkhipovFasterAlgorithmsPacking2025} once on the contracted graph to give a largest union of $k$ forests, but it may miss some arcs of $\Ix$. We repair it by adding these arcs back one at a time, and each one replaces one of the chosen terminals (\Cref{sec:compatible-set-selection}). Since our paths are matroid-level, executing them does not maintain the explicit forest decomposition. So we will need to rebuild the forests once at the end of the phase, with the same algorithm. These two steps take $O(k^3n\log(kn))$ time per phase.

\paragraph{Progress (\Cref{sec:batch-optimality-progress}).}
Let $\defi_I(V)=k(n-1)-|I|$ be the number of missing arcs. After a phase, we show that every deficient tight block of the new set contains the head of an executed terminal. Otherwise, the arcs inside the block are unchanged, and the terminal of an old proposal inside it could still be added, which contradicts maximality. The number of missing arcs drops by $|A'|$, and at most $|A'|$ tight blocks of the new set are deficient, each missing at most $k$ arcs. Hence $\defi_I(V)$ drops by at least a $\frac{1}{k+1}$ fraction, and $O(k\log(kn))$ phases suffice (\Cref{lem:batch-progress}).

\paragraph{Compressing the exchange graph (\Cref{sec:search-graph-compression}).}
Since there are $O(k\log(kn))$ phases, it suffices to make each phase near-linear. The remaining cost is the shortest-path computation that finds all proposals: the exchange graph can have $\Theta(kmn)$ arcs, and its arc weights can be negative.

Gabow finds his fine-grained augmenting paths by a reachability search in a \emph{fine-grained search graph}: like the exchange graph, its vertices are the arcs of $G$, and it has a directed edge $f\to e$ for $f\in I$ and $e\notin I$ with the same head. Instead of the matroid-level exchange edges, it has a directed edge $x\to y$ for every \emph{forest swap}: $y$ lies in a forest $F_j$ with $x\notin F_j$, on the $F_j$-path between the ends of $x$, so $x$ can replace $y$ in $F_j$. Each path in this graph projects to a matroid-level augmenting path with the same cost (\Cref{lem:fine-grained-projection}), so we can also use it to find our paths.

Gabow avoids building this graph by visiting the elements in an order he calls cyclic scanning~\cite[Section~2.3]{gabowMatroidApproachFinding1995}. This order lets him list the unvisited neighbors of the visited set quickly. In the weighted case, we need shortest paths, so we cannot visit the elements in this order.

We therefore build this graph explicitly and make it small. We add one auxiliary vertex per head for the exchanges in the partition matroid, and compress the tree paths with a sparse-table-like structure (following the approach of~\cite{saranurakIncrementalDirectedMinimum2026}), so that each element needs at most four arcs per forest (\Cref{sec:tree-path-compression}). We further show that for each element, it suffices to keep its swaps into a single forest, saving a factor of $k$ in the number of scan arcs (\Cref{app:cyclic-scans}).
The resulting graph has $O(kn\log n)$ vertices and $O(m+kn\log n)$ arcs and preserves shortest paths.

With black-box negative-weight shortest-path algorithms for the proposals, this already gives a weaker version of \Cref{thm:main}: for fixed $k$, near-linear time with logarithmic cost dependence for integer costs, and almost-linear time for real costs (\Cref{rem:black-box-shortest-paths}).

\paragraph{Strongly polynomial algorithm via vertex prices (\Cref{sec:vertex-prices}).}

To get better runtime, we maintain a potential on the search graph to make all arc weights nonnegative. We maintain a price $h(v)$ for each vertex $v$, and assign each arc $e$ the adjusted cost $a(e)=c(e)-h(\head(e))$. Our invariant is that $I$ is a minimum-cost independent set of $\mathcal G^k$ under the adjusted costs (with no indegree constraints).

Under these costs, a $\mathcal G^k$-exchange arc in the exchange graph has weight $a(e)-a(f)$, which is nonnegative by the optimality of $I$. For an augmenting path from $u$, the prices cancel except for $h(u)$, so its adjusted cost is its original cost minus $h(u)$. Thus the shortest augmenting paths from each vertex remain shortest.

To preserve the invariant after augmentation, we update the vertex prices so that the old set remains optimal and every shortest proposal has zero adjusted cost. Any compatible batch then preserves optimality for the new adjusted costs, so the next phase starts with the same invariant.

However, the compressed graph represents each exchange by several arcs, whose individual weights may still be negative. We extend the vertex prices to a feasible potential on the compressed graph by propagating adjusted-cost bounds along the scan arcs in linear time. This lets us find all proposals with one Dijkstra computation per phase, giving the deterministic, strongly polynomial algorithm of \Cref{thm:main}.

\paragraph{Indegree-bounded forest packing (\Cref{sec:indegree-bounded-forest-packing}).}
If the indegree of each vertex $v$ may be at most $\bar b_v\le k$, we add a new root with $k-\bar b_v$ very cheap and $\bar b_v$ very expensive arcs into each $v$. The cheap arcs use up the capacity that the bound removes, the expensive ones mark unused slots, and the original arcs of an optimal packing form an optimal solution (\Cref{thm:indegree-bounded-forest-packing}).

\paragraph{Undirected forest packing (\Cref{sec:undirected-forest-packing}).}
For undirected graphs, we follow the outer loop of the unweighted algorithm of Arkhipov and Kolmogorov~\cite{arkhipovFasterAlgorithmsPacking2025}. Each round finds a cheapest maximum union of $k$ pseudoforests (graphs in which every component has at most one cycle) with min-cost flow, orients it with indegree at most $k$, finds a cheapest maximum union of $k$ forests inside it with the indegree-bounded algorithm above, and contracts the resulting tight blocks.

We prove a two-stage greedy lemma: the greedy union of $k$ forests inside the greedy union of $k$ pseudoforests is contained in the greedy optimal solution (\Cref{lem:two-stage-greedy}). Hence every contraction fixes only edges of an optimal solution. The progress argument is the same as in~\cite{arkhipovFasterAlgorithmsPacking2025}.

\section{Preliminaries}
\label{sec:prelim}

\subsection{Notation and Matroid Terminology}
\label{sec:notation}

We use the graph notation from the introduction, and regard parallel arcs as distinct elements. For arborescence packing, we delete loops and arcs entering $r$, and assume $n\ge2$.

We write $\log x=\max\{1,\log_2x\}$. We do not assume that $k$ is polynomially bounded in $n$, so some running times contain $\log(kn)$. Exact arithmetic operations and comparisons take constant time.

For a set $Y$ and an element $y$, we write $Y+y=Y\cup\{y\}$ and $Y-y=Y\setminus\{y\}$. We write $Y\oplus Z$ for the symmetric difference of sets $Y,Z$. For $X\subseteq E$, let $c(X)=\sum_{e\in X}c(e)$. For an arc $e$, $\head(e)$ denotes its head. For $S\subseteq V$, $X[S]$ is the set of arcs of $X$ with both ends in $S$. We write $\degin_X(v)$ for the number of arcs of $X$ entering $v$, and $\degin_X(S)$ for the number entering $S$ from $V\setminus S$. For $U\subseteq V$, $\one_U$ is its indicator vector and $\one_u=\one_{\{u\}}$.

We use standard matroid terminology~\cite{schrijverCombinatorialOptimizationPolyhedra2004}, writing $\rk_M$ and $\cl_M$ for the rank and closure operators of a matroid $M$. If $I$ is independent and $I+e$ is not, then $I+e$ contains a unique circuit, the \emph{fundamental circuit} of $e$; for $f\in I$, the set $I-f+e$ is independent iff $f$ lies in this circuit.

\subsection{Unions of Forests and Tight Blocks}
\label{sec:forests-tight-blocks}

Let $\mathcal G$ be the graphic matroid of the undirected multigraph underlying $G$, and let $\mathcal G^k$ be the union of $k$ copies of $\mathcal G$; its independent sets are the unions of $k$ forests. By Nash-Williams~\cite{nashwilliamsDecompositionFiniteGraphs1964}, $I\subseteq E$ is independent in $\mathcal G^k$ iff
\begin{equation}\label{eq:forest-density}
	|I[S]|\le k(|S|-1)\qquad\text{for every nonempty }S\subseteq V.
\end{equation}
A \emph{forest decomposition} of $I$ is a partition $I=F_1\dotcup\cdots\dotcup F_k$ into forests. An element $e\notin I$ is \emph{free} if $I+e$ is independent.

When $\degin_I(r)=0$ and $\degin_I(v)\le k$ for every $v\ne r$, the \emph{deficiency} of a nonroot vertex $v$ is $\defi_I(v)=k-\degin_I(v)$. Set $\defi_I(r)=0$ and $\defi_I(S)=\sum_{v\in S}\defi_I(v)$.

\begin{lemma}[Matroid formulation; Edmonds~\cite{edmondsEdgeDisjointBranchings1973}]
	\label{lem:arborescence-formulation}
	$I\subseteq E$ is the arc set of $k$ arc-disjoint $r$-arborescences iff $I$ is independent in $\mathcal G^k$, $\degin_I(r)=0$, and $\degin_I(v)=k$ for every $v\ne r$.
\end{lemma}
\begin{proof}
	One direction is clear. Conversely, if $I$ is independent with these indegrees, then every nonempty $S\subseteq V-r$ has $\degin_I(S)=k|S|-|I[S]|\ge k$ by~\eqref{eq:forest-density}, and Edmonds' theorem~\cite{edmondsEdgeDisjointBranchings1973} gives $k$ arc-disjoint $r$-arborescences in $I$. They use all $k(n-1)=|I|$ arcs.
\end{proof}

\paragraph*{Tight blocks.}
Let $I$ be independent. A nonempty set $T\subseteq V$ is \emph{$I$-tight} if $|I[T]|=k(|T|-1)$; equivalently, every forest of a forest decomposition of $I$, restricted to $T$, is a spanning tree of $T$. Intersecting $I$-tight sets have an $I$-tight union, so the maximal $I$-tight sets partition $V$. We call this partition $\cT(I)$ and its members the \emph{$I$-tight blocks}. A vertex or tight block is \emph{deficient} if its deficiency is positive. An arc is \emph{internal} if both its ends lie in one block, and \emph{crossing} otherwise.

\begin{lemma}[Free elements]
	\label{lem:free-elements}
	An element $e\notin I$ is free iff it is crossing. If $e$ is internal to a block $T$, then its fundamental circuit lies in $I[T]+e$.
\end{lemma}
\begin{proof}
	Adding $e$ increases $|I[S]|$ only for sets $S$ containing both ends of $e$, so~\eqref{eq:forest-density} fails for $I+e$ iff some tight set contains both ends. If $e$ is internal to $T$, then $I[T]+e$ is dependent.
\end{proof}

\begin{lemma}[Deficiency of a tight block]
	\label{lem:block-deficiency}
	Let $I$ have no arc entering $r$ and indegree at most $k$ at every other vertex, and let $T$ be $I$-tight. Then $\defi_I(T)=k-\degin_I(T)\le k$ if $r\notin T$, and $\defi_I(T)=0$ if $r\in T$.
\end{lemma}
\begin{proof}
	The heads in $T$ receive $|I[T]|+\degin_I(T)=k(|T|-1)+\degin_I(T)$ arcs, and their capacity is $k|T|$ if $r\notin T$ and $k(|T|-1)$ if $r\in T$.
\end{proof}

\begin{lemma}[Computing tight blocks~{\cite[Theorem~5.4]{gabowForestsFramesGames1992}}; see {\cite[Lemma~4.4]{arkhipovFasterAlgorithmsPacking2025}}]
	\label{lem:tight-block-computation}
	Given a forest decomposition of $I$, the partition $\cT(I)$ can be computed deterministically in $O(kn\log n)$ time.
\end{lemma}

The tight blocks are the vertex sets of the connected components of the top clump of $I$, together with the remaining singletons; thus the cited top-clump algorithm computes $\cT(I)$.\footnote{A \emph{clump} is a set $L\subseteq I$ with $|L|=k\rk_{\mathcal G}(L)$; the \emph{top clump} is the unique inclusion-maximal clump~\cite[Definition~4.3]{arkhipovFasterAlgorithmsPacking2025}. Every tight set induces a connected clump, except for singletons, and hence lies in one component of the top clump. Conversely, equality in the clump size bound and~\eqref{eq:forest-density} force every component of a clump to be tight.}

\subsection{Weighted Matroid Intersection}
\label{sec:weighted-matroid-intersection}

Let $M_1,M_2$ be matroids on $E$, and let $c:E\to\mathbb R$ be costs. A set is \emph{common independent} if it is independent in both matroids. A common independent set $I$ is \emph{extreme} if $c(I)\le c(X)$ for every common independent $X$ with $|X|=|I|$. Following~\cite[Section~41.2]{schrijverCombinatorialOptimizationPolyhedra2004}, let
\[
	X_1(I)=\{e\notin I: I+e\text{ is independent in }M_1\},\qquad
	X_2(I)=\{e\notin I: I+e\text{ is independent in }M_2\}.
\]

\paragraph*{Exchange graph.}
The \emph{exchange graph} $D(I)=D_{M_1,M_2}(I)$ has vertex set $E$. For $e\notin I$ and $f\in I$, it has
\begin{itemize}
	\item an arc $e\to f$ of weight $c(e)-c(f)$ if $I-f+e$ is independent in $M_1$, and
	\item an arc $f\to e$ of weight $0$ if $I-f+e$ is independent in $M_2$.
\end{itemize}

\paragraph*{Augmenting paths.}
An \emph{augmenting path} is a simple path $P=(e_0,f_0,e_1,f_1,\ldots,f_{q-1},e_q)$ in $D(I)$ with $e_0\in X_2(I)$ and $e_q\in X_1(I)$. We identify $P$ with its element set. Its \emph{cost} is
\[
	\ell(P)=c(I\oplus P)-c(I)=\sum_{j=0}^q c(e_j)-\sum_{j=0}^{q-1}c(f_j),
\]
and its \emph{exchange count} is $q(P)=|P\cap I|=q$, counting only the $M_1$-exchanges. With the arc weights of $D(I)$, the cost of $P$ equals the cost of its last element plus the sum of its arc weights. Throughout, a \emph{shortest augmenting path} minimizes $\ell(P)$, with ties broken by minimizing $q(P)$.

\begin{theorem}[{\cite[Theorems~41.2--41.3 and~41.5--41.6]{schrijverCombinatorialOptimizationPolyhedra2004}}]
	\label{thm:weighted-matroid-intersection}
	A common independent set $I$ has maximum cardinality iff $D(I)$ has no augmenting path. It is extreme iff $D(I)$ has no negative cycle. If $I$ is extreme and $P$ is a shortest augmenting path, then $I\oplus P$ is an extreme common independent set of size $|I|+1$.
\end{theorem}

\subsection{Algorithmic Tools}
\label{sec:algorithmic-tools}

\begin{lemma}[Unweighted forest packing; {\cite[Theorem~1.1 and Appendix~A.5]{arkhipovFasterAlgorithmsPacking2025}}]
	\label{lem:ak-forest-packing}
	Let $H$ be a digraph with $n_H$ vertices and $m_H$ arcs. The algorithm of Arkhipov and Kolmogorov computes a maximum-cardinality union of $k$ forests in $H$ with indegree at most $k$ at every vertex, together with a forest decomposition, deterministically in $O(k^2(m_H+n_H)\log(kn_H))$ time. If every vertex of $H$ has indegree at most $k$, the output is a basis of $\mathcal G^k(H)$.
\end{lemma}

\begin{lemma}[Arborescence extraction; {\cite[Theorem~1.4]{bhalgatFastEdgeSplitting2008}}]
	\label{lem:arborescence-extraction}
	If an unweighted digraph contains $k$ arc-disjoint $r$-arborescences, then the algorithm of Bhalgat, Hariharan, Kavitha, and Panigrahi finds $k$ such arborescences deterministically in $O(km\log n+k^4n\log^2n)$ time.
\end{lemma}

\section{Min-Cost \texorpdfstring{$k$}{k}-Arborescence Packing}\label{sec:arborescence-packing}

This section presents our algorithm, following the plan of \Cref{sec:overview}.

\Cref{sec:relax-and-augment} works with an arbitrary matroid intersected with a partition matroid. It shows that a shortest augmenting path raises one class count and keeps the set cheapest for its class counts, and that raising several counts together costs at least the sum of the single-count costs (\Cref{thm:relaxation}). We defer the proof to \Cref{sec:relaxation-proof}.

\Cref{sec:batching-framework} specializes to $\mathcal G^k$ and builds the batching algorithm from this theorem. In each phase, every deficient tight block proposes one shortest augmenting path, which stays inside the block except for its last arc (\Cref{sec:localized-proposals}). \Cref{sec:contracted-compatibility} then gives a criterion, based on contracting the tight blocks, for when a set of proposals can be executed together while keeping $I$ independent. \Cref{sec:batch-optimality-progress} shows that executing any maximal such set keeps $I$ optimal and reduces the number of missing arcs by at least a $\frac{1}{k+1}$ fraction, so $O(k\log(kn))$ phases suffice.

\Cref{sec:warmup-complexity} reduces the total running time to two steps per phase: finding the proposals, and choosing a largest compatible set. \Cref{sec:search-graph-compression,sec:compatible-set-selection} implement these two steps. The first replaces the exchange graph by a search graph of near-linear size with the same shortest-path distances. The second finds a largest compatible set with one call to the forest-packing algorithm of Arkhipov and Kolmogorov~\cite{arkhipovFasterAlgorithmsPacking2025}, followed by a repair step. With black-box negative-weight shortest paths, this already gives a weaker version of our main result (\Cref{rem:black-box-shortest-paths}).

\subsection{Relax and Augment}
\label{sec:relax-and-augment}

In this subsection, we establish the optimality properties used by our batching algorithm. We first show how to increase one indegree optimally, and then give a lower bound on the cost of increasing several indegrees together. We work with an arbitrary matroid $M$ intersected with a partition matroid, where class counts play the role of indegree bounds.

\paragraph*{Setup.}
Let $M$ be a matroid on $E$, and let $E$ be partitioned into classes $E_v$, $v\in V$. For an integer vector $\vb\ge\mathbf 0$, let $N_{\vb}$ be the partition matroid with capacity $b_v$ on $E_v$. To \emph{relax} a class $u$ means to raise $b_u$ by one. Let
\[
	\OPT(\vb)=\min\{c(X): X\text{ is independent in }M\text{ and }|X\cap E_v|=b_v\text{ for all }v\},
\]
with $\OPT(\vb)=+\infty$ if no such $X$ exists. An independent set $I$ of $M$ with class counts $\vb$ is extreme for $M$ and $N_{\vb}$ iff $c(I)=\OPT(\vb)$.

\paragraph*{Augmenting paths.}
Fix an independent set $I$ of $M$ with class counts $\vb$, and let $D(I)=D_{M,N_{\vb}}(I)$ be the exchange graph.

For elements $e\notin I$ and $f\in I$, the graph has an arc $e\to f$ iff $I-f+e$ is independent in $M$, and an arc $f\to e$ iff $I-f+e$ is independent in $N_{\vb}$. Since every class is full, the condition for $N_{\vb}$ holds exactly when $e$ and $f$ belong to the same class.

An \emph{augmenting path from class $u$} is a simple path $P=(e_0,f_0,\ldots,f_{q-1},e_q)$ in $D(I)$ with $e_0\in E_u\setminus I$ and $e_q\in X_1(I)$ (recall that $X_1(I)=\{e\in E\setminus I: I+e\text{ is independent in }M\}$). If $I\oplus P$ is independent, its class counts are $\vb+\one_u$, since $e_0$ enters class $u$ and each $f_j$ is replaced by $e_{j+1}$ from the same class. We define $\ell(P)$ and $q(P)$ as in \Cref{sec:weighted-matroid-intersection}, and a shortest augmenting path from class $u$ minimizes cost among those paths, breaking ties by the number of exchanges. Let $\Delta_u=\OPT(\vb+\one_u)-\OPT(\vb)$.

\begin{remark}
	\label{rem:current-capacities}
	Note that an augmenting path from class $u$ is a path in $D_{M,N_{\vb}}(I)$, the exchange graph for the \emph{current} capacities $\vb$, and not in $D_{M,N_{\vb+\one_u}}(I)$. This differs from the standard algorithm behind \Cref{thm:weighted-matroid-intersection}. Every class is full for $N_{\vb}$, so $X_2(I)=\emptyset$, and we start at $E_u\setminus I$ instead.

	Searching in $D_{M,N_{\vb+\one_u}}(I)$ would be wrong. In general, $I$ is not extreme for $M$ and $N_{\vb+\one_u}$: the common independent sets of size $|I|$ may then have class counts $\vb+\one_u-\one_w$ for any $w$, and some of them may be cheaper than $I$. By \Cref{thm:weighted-matroid-intersection}, $D_{M,N_{\vb+\one_u}}(I)$ may then have negative cycles, and a shortest augmenting path need not exist.
\end{remark}

\begin{theorem}[Relaxation theorem]
	\label{thm:relaxation}
	Let $I$ be extreme with class counts $\vb$.
	\begin{enumerate}[label=(\alph*)]
		\item If $\OPT(\vb+\one_u)<+\infty$, then an augmenting path from class $u$ exists. For every shortest augmenting path $P$ from class $u$, the set $I\oplus P$ is independent with class counts $\vb+\one_u$, and $\ell(P)=\Delta_u$.
		\item For every $U\subseteq V$, we have $\OPT(\vb+\one_U)\ge \OPT(\vb)+\sum_{u\in U}\Delta_u$.\footnote{This lower bound also appears to follow from the aggregation results in~\cite[Proposition~4.6(2)]{murotaBasicOperationsRelated2020}.}
	\end{enumerate}
\end{theorem}

Part~(a) shows that a shortest augmenting path relaxes one
class optimally. Part~(b) implies that executing several
such paths together is also optimal, provided that the paths
are element-disjoint and the resulting set is independent.
We give an elementary proof in \Cref{sec:relaxation-proof}.

The next lemma shows that a shortest augmenting path stops at its first free element.

\begin{lemma}[Only the Last Element Is Free]
	\label{lem:no-nonterminal-free-elements}
	Let $I$ be extreme, and let $P=(e_0,f_0,\ldots,f_{q-1},e_q)$ be a shortest augmenting path from class $u$. Then $e_0,\ldots,e_{q-1}\notin X_1(I)$.
\end{lemma}
\begin{proof}
	Suppose $e_i\in X_1(I)$ for some $i<q$. The prefix $P'=(e_0,f_0,\ldots,e_i)$ is an augmenting path from class $u$ with fewer exchanges, so cost minimality and the tie-breaking rule give $\ell(P')>\ell(P)$. Since $I+e_q$ is independent in $M$, so is $I-f_i+e_q$. Hence $D(I)$ has the arc $e_q\to f_i$, and $e_q\to f_i\to e_{i+1}\to f_{i+1}\to\cdots\to f_{q-1}\to e_q$ is a cycle in $D(I)$ of weight
	\[
		\sum_{j=i+1}^{q}c(e_j)-\sum_{j=i}^{q-1}c(f_j)=\ell(P)-\ell(P')<0,
	\]
	which is a negative cycle, contradicting \Cref{thm:weighted-matroid-intersection}.
\end{proof}

\paragraph*{The graph setting.}
We use $M=\mathcal G^k$, with $E_v$ the set of arcs entering $v$. Then class counts are indegrees, and $X_1(I)$ is the set of free arcs. For $e\notin I$ and $f\in I$, the exchange graph $D(I)$ has
\begin{itemize}
	\item an arc $e\to f$ of weight $c(e)-c(f)$ if $e$ is free or $f$ lies on the fundamental circuit of $e$, and
	\item an arc $f\to e$ of weight $0$ if $\head(e)=\head(f)$.
\end{itemize}
An augmenting path from class $u$ starts at an arc $e\notin I$ entering $u$ and ends at a free arc.

\subsection{Batching Framework}\label{sec:batching-framework}

By \Cref{thm:relaxation}, element-disjoint shortest augmenting paths from distinct vertices preserve optimality if applying them together preserves independence. We use tight blocks to select such paths: each deficient block proposes one path, and we choose a compatible subset of them. We then show that this selection guarantees progress in every phase.

We maintain an independent set $I$ that has minimum cost among all independent sets with the same indegree vector $\vb=\degin_I$. We also maintain a forest decomposition of $I$. Initially, $I=\varnothing$. Throughout, no arc of $I$ enters $r$, and every other vertex has indegree at most $k$. In \Cref{sec:vertex-prices}, we maintain a stronger optimality certificate to implement the shortest-path searches with Dijkstra's algorithm.

\subsubsection{Locality of Augmenting Paths}\label{sec:localized-proposals}
We use the exchange graph $D(I)$ and the augmenting paths of \Cref{sec:relax-and-augment}, for $M=\mathcal G^k$.

\begin{lemma}[Localization to a tight block]\label{lem:proposal-localization}
	Let $u$ lie in a tight block $T\in\cT(I)$, and let $P$ be a shortest augmenting path from $u$. All elements of $P$ except the last are internal to $T$, and the last element is a free arc entering $T$; we call it the \emph{terminal} of $P$.

	Moreover, $I'=I\oplus P$ is independent and satisfies
	\begin{equation}\label{eq:localization-count}
		\degin_{I'}=\degin_I+\one_u,
		\qquad |I'[T]|=|I[T]|=k(|T|-1).
	\end{equation}
\end{lemma}
\begin{proof}
	By \Cref{lem:no-nonterminal-free-elements}, no element of $P$ before its last one is free. By \Cref{lem:free-elements}, a nonfree element $e\notin I$ with head in $T$ is internal to $T$, and its fundamental circuit lies in $I[T]$ plus that element. Hence each exchange $e_j\to f_j$ stays in $T$, and so does the next step $f_j\to e_{j+1}$, which goes to an arc with the same head. The final element is free, so its tail is outside $T$. The same localization argument applies to any augmenting path with no free element before its last one.

	The augmenting-path criterion and \Cref{thm:relaxation}(a) give independence of $I'$ and the stated indegree vector. Before its terminal, $P$ removes as many elements of $I[T]$ as it adds from $E[T]\setminus I$, giving the second equality in~\eqref{eq:localization-count}. Thus $T$ remains tight after augmentation.
\end{proof}

In each deficient block $T$, relax one deficient vertex $u_T$, and let the \emph{proposal} $P_T$ be a shortest augmenting path from $u_T$, with terminal $e_T$. Such a path must exist: assuming a packing exists, deleting arcs from it gives an independent set with indegree vector $\vb+\one_{u_T}$, so $\OPT(\vb+\one_{u_T})<+\infty$ and a proposal exists by \Cref{thm:relaxation}(a). Therefore, if some deficient block has no proposal, we report that no packing exists.

Different proposals are element-disjoint: their internal elements belong to disjoint blocks, and their crossing terminals have heads in those blocks and are thus distinct.

Let $\mathsf S(N,M)$ denote the running time of the shortest-path subroutine on a graph with $N$ vertices and $M$ arcs. For now, we compute the proposals using $D(I)$. But recall that the cost of an augmenting path $P$ equals the sum of its arc weights plus the cost of its last element. We therefore add a sink $z$ and an arc $e\to z$ of weight $c(e)$ for each free $e$.

To compute the shortest-path distance $d(e)$ from each element $e$ to $z$, we simply run a reverse SSSP computation from $z$. Then the minimum cost from a class $v$ is $\min_{e\in E_v\setminus I}d(e)$. We defer the efficient implementation to \Cref{sec:search-graph-compression}.

\subsubsection{Checking Compatibility by Contracting Tight Blocks}\label{sec:contracted-compatibility}

By \Cref{lem:proposal-localization},
each proposal replaces $I[T]$ by another basis of $E[T]$ in
$\mathcal G^k$ and adds its terminal. Intuitively, the choice
of the internal basis should not affect whether the proposals
can be applied together.

We therefore contract each block of $\cT(I)$ to a vertex. Let $\Ix$ be the set of arcs in $I$ whose endpoints lie in different blocks, and let $A=\{e_T\}$ be the
set of terminals. We check compatibility in the contracted graph with vertex set $\cT(I)$ and arc set
\[
	H=\Ix\dotcup A.
\]
Independence is still taken in the underlying undirected graph.

The following lemma shows that the internal replacements do not
affect compatibility.

\begin{lemma}[Contraction and lifting]\label{lem:contraction-lifting}
	For every set $X$ of arcs crossing the old blocks, the following are equivalent:
	\begin{enumerate}[label=(\alph*)]
		\item $I\cup X$ is independent in the original graph;
		\item $\Ix\cup X$ is a union of $k$ forests in the contracted graph.
	\end{enumerate}
	The equivalence is unchanged if any internal basis $I[T]$ is replaced by another basis of $E[T]$.
\end{lemma}
\begin{proof}
	In a forest decomposition of an independent set containing the internal bases, the $k(|T|-1)$ arcs inside $T$ force each forest restricted to $T$ to be a spanning tree. Contract these trees in every forest to obtain a forest decomposition of the contracted graph.

	Conversely, combine any forest decomposition of the contracted graph with a decomposition of each internal basis: adding its $i$th tree inside each contracted vertex to the $i$th forest of the contracted graph gives a forest.
\end{proof}

Formally, we call $A'\subseteq A$ \emph{compatible} if $\Ix\cup A'$ is independent in the contracted graph. We find a largest compatible $A'$ and execute the proposals whose terminals lie in $A'$. \Cref{lem:contraction-lifting} shows that executing them together gives an independent set.

\begin{lemma}[Indegree bounds after contraction]\label{lem:contracted-indegrees}
	Every vertex of the contracted graph has indegree at most $k$. If $t=|\cT(I)|$, then $|\Ix|+|A|\le k(t-1)$.
\end{lemma}
\begin{proof}
	For a nonroot tight block $T$ of $I$, \Cref{lem:block-deficiency} gives $\degin_{\Ix}(T)=\degin_{I}(T)=k-\defi_I(T)$. As we find a proposal for $T$ iff $\defi_I(T)>0$, in such a case the terminal raises $\degin_{\Ix}(T)$ by one, and we still have $\degin_{\Ix\cup A}(T)\le k$.

	The root block has no entering arc of $I$ and no proposal. Summing the bounds over the other $t-1$ blocks proves the claim.
\end{proof}

\paragraph{Keeping old crossing arcs.}
Because the indegree in the contracted graph is at most $k$, satisfying the requirements of \Cref{lem:ak-forest-packing}, the idea is to use this algorithm to compute a maximum-cardinality forest packing $B$ of $H$, but this need not contain $\Ix$. The proposals leave all old crossing arcs unchanged, so the chosen terminals must be independent together with $\Ix$. What we actually need is an extension of $\Ix$ to a basis of $H$. This can be done by repairing $B$, as described in \Cref{sec:compatible-set-selection}. For now, we denote the running time of this step by $\Sigma$.

We summarize the procedure described above as the following pseudocode:
\begin{algorithm}[H]
	\caption{Packing by selection on the contracted graph}\label{alg:batching-framework}
	\begin{algorithmic}[1]
		\Require digraph $G$, root $r$, integer $k\ge1$, costs $c:E\to\mathbb R$
		\State delete loops and arcs entering $r$; if $n=1$ return the empty packing
		\State $I\gets\varnothing$ with empty forests
		\While{$|I|<k(n-1)$}
		\State compute $\cT(I)$; choose a deficient vertex $u_T$ in each deficient block
		\State for each $u_T$, find a shortest augmenting path $P_T$ from $u_T$ in $D(I)$; if none exists, return ``infeasible''
		\State contract the old blocks; form the arc set $H=\Ix\dotcup A$ with $A=\{\text{terminal of }P_T\}$
		\State extend $\Ix$ to a basis $K$ of $H$
		\State $I\gets I\oplus\displaystyle\bigcup_{T:e_T\in K\setminus \Ix}P_T$ \Comment{execute the chosen proposals}\label{line:execute-proposals}
		\State compute a forest decomposition of $I$ using \Cref{lem:ak-forest-packing}
		\EndWhile
		\State partition $I$ into $k$ arborescences using \Cref{lem:arborescence-extraction}; return them
	\end{algorithmic}
\end{algorithm}

The augmentation in \cref{line:execute-proposals} preserves independence but does not explicitly maintain a forest decomposition. Since the updated $I$ satisfies the indegree bounds, we simply compute a fresh forest decomposition using \Cref{lem:ak-forest-packing} on $(V,I)$.

\subsubsection{Optimality and Progress}\label{sec:batch-optimality-progress}

\begin{lemma}[A compatible batch is optimal]\label{lem:batch-optimality}
	Let $A'$ be any compatible set of terminals, let $U=\{u_T:e_T\in A'\}$, and set
	\[
		I'=I\oplus\bigcup_{T:e_T\in A'}P_T.
	\]
	Then $I'$ is a minimum-cost independent set with indegree vector $\vb+\one_U$.
\end{lemma}
\begin{proof}
	By \Cref{lem:proposal-localization,lem:contraction-lifting}, compatibility of $A'$ implies that $I'$ is independent. By \Cref{lem:proposal-localization} and element-disjointness of the paths, $\degin_{I'}=\vb+\one_U$.
	By \Cref{thm:relaxation}(a), we have $\ell(P_T)=\Delta_{u_T}\coloneqq \OPT(\vb+\one_{u_T})-\OPT(\vb)$ for each $T$. Then, by element-disjointness and \Cref{thm:relaxation}(b), we have
	\begin{align*}
		\OPT(\vb+\one_U)\le c(I')
		 & =c(I)+\sum_{T:e_T\in A'}\ell(P_T) \\
		 & =\OPT(\vb)+\sum_{u\in U}\Delta_u
		\le\OPT(\vb+\one_U).
	\end{align*}
	Thus equality holds.
\end{proof}

\begin{lemma}[Geometric progress]\label{lem:batch-progress}
	Suppose $\defi_I(V)>0$, and let $A'$ be a maximal compatible set of terminals. Then the resulting $I'$ satisfies $\defi_{I'}(V)=\defi_I(V)-|A'|\le k|A'|$.

	Consequently, total deficiency decreases by a factor of $k/(k+1)$ in each phase, and there are $O(k\log(kn))$ phases.
\end{lemma}
\begin{proof}

	By~\eqref{eq:localization-count}, every old block remains tight in $I'$, so each new tight block is a union of old blocks.

	We claim that every deficient new block $S$ contains the \emph{head} of a chosen terminal. Suppose not. Then no old block inside $S$ executes a proposal, so $I'[S]=I[S]$ and the indegrees in $S$ are unchanged.

	Since $|I'[S]|=k(|S|-1)$ and $I[S]=I'[S]$, the set $S$ is $I$-tight as well, so $S$ itself is a single old tight block. Since $S$ was already deficient in $I$, it must have proposed $e_S$, whose tail lies outside $S$.

	However, since $S$ is also an $I'$-tight block, the arc $e_S$ is still a crossing arc for $I'$, and thus $I'+e_S$ is independent by \Cref{lem:free-elements}. By \Cref{lem:contraction-lifting}, this means that $A'+e_S$ is compatible, contradicting the maximality of $A'$.

	Hence there are at most $|A'|$ deficient new blocks, each of deficiency at most $k$. This proves $\defi_I(V)-|A'|\le k|A'|$. The deficiency thus shrinks by a factor $k/(k+1)$ per phase. It starts at $k(n-1)$, which gives the phase bound.
\end{proof}

The proof uses only the fact that $A'$ is maximal.

\subsubsection{Warm-Up Complexity}\label{sec:warmup-complexity}
We first bound the running time in terms of the cost of finding the proposals and the cost of compatible selection. This makes clear what we need to improve.

\begin{lemma}[Parametrized running time]\label{lem:parametric-running-time}
	Suppose each phase finds all proposals in $O(\mathsf S(N,M))$ time and a largest compatible set of terminals in time $\Sigma$, where $\mathsf S(N,M)\ge N+M$. Then the entire algorithm runs in
	\begin{equation}\label{eq:parametric-running-time}
		O\Bigl(k\log(kn)\cdot\bigl(\mathsf S(N,M)+\Sigma+k^3n\log(kn)\bigr)+km\log n\Bigr)
	\end{equation}
	time.
\end{lemma}
\begin{proof}
	By \Cref{lem:batch-progress} there are $O(k\log(kn))$ phases. In one phase,  computing the tight blocks of $I$ takes $O(kn\log n)$ time by \Cref{lem:tight-block-computation}, since $|I|\le kn$. Finding the proposals costs $O(\mathsf S(N,M))$. Building the contracted graph costs $O(kn)$ by \Cref{lem:contracted-indegrees}, and finding a largest compatible set of terminals costs $\Sigma$. Applying the chosen proposals costs $O(kn)$, since their total length is at most $2|I|+|A'|$ by element-disjointness. Rebuilding the forest decomposition of $I$ costs $O(k^3n\log(kn))$ by \Cref{lem:ak-forest-packing}, as $|I|\le kn$. An infeasible input stops in the first phase in which some deficient block has no proposal. The final extraction costs $O(km\log n+k^4n\log^2n)$ by \Cref{lem:arborescence-extraction}, and $k^4n\log^2n$ is within the first term.
\end{proof}

For fixed $k$, \eqref{eq:parametric-running-time} is near-linear if each phase finds all proposals and a largest compatible set of terminals in near-linear time. For the search step, there are two remaining issues.
\begin{itemize}
	\item \emph{Size of the search graph.} The exchange graph $D(I)$ has $m$ vertices and $O(kmn)$ arcs, since an element of $E\setminus I$ can have an arc to each of the at most $kn$ elements of $I$. Building it also requires the fundamental circuits of the nonfree elements in $E\setminus I$.
	\item \emph{Negative-weight shortest paths.} The graph has negative arc weights, so we cannot directly use Dijkstra's algorithm.
\end{itemize}

\begin{remark}[Implementations using black-box shortest paths]
	\label{rem:black-box-shortest-paths}
	On a graph with $N$ vertices, $M$ arcs, and no negative cycle, exact SSSP takes $O(M\log^8N\log(NC))$ time deterministically for integer weights in $[-C,C]$, where $C\ge1$~\cite[Theorem~1.1]{haeuplerDeterministicNegativeWeight2026}, or $M^{1+o(1)}$ time with high probability for real weights using a Las Vegas algorithm~\cite{hairBellmanFordAlmostLinear2026}.
	We combine these SSSP running times with the compression in \Cref{sec:search-graph-compression} and the compatible-set computation in \Cref{sec:compatible-set-selection}. Substituting into~\eqref{eq:parametric-running-time}, we obtain a deterministic algorithm for integer costs with $|c(e)|\le C$ running in
	\[
		O\bigl(k(m+kn\log n)\log^9(kn)\log(knC)+k^4n\log^2(kn)\bigr)
	\]
	time, and a Las Vegas algorithm for arbitrary real costs running in
	\[
		O\bigl(k\log(kn)(m+kn\log n)^{1+o(1)}+k^4n\log^2(kn)\bigr)
	\]
	time with high probability. The compressed arcs have weights $0$ or $\pm c(e)$, so compression preserves the integer-weight bound. The randomized integer-weight algorithm of~\cite{bernsteinNegativeWeightSingleSource2022} also achieves the first running time with high probability.
\end{remark}

In \Cref{sec:vertex-prices}, we construct a feasible potential function and use Dijkstra's algorithm to obtain the strongly polynomial running time stated in \Cref{thm:main}.

\subsection{Compressing the Exchange Graph}\label{sec:search-graph-compression}

Throughout this subsection, fix the current set $I$ and its forest decomposition $I=F_1\dotcup\cdots\dotcup F_k$. We suppress this dependence in the notation for the search graphs, writing $D$ for $D(I)$.

We find the matroid-level augmenting paths using a fine-grained search graph based on forest swaps, similar to Gabow's search graph~\cite{gabowMatroidApproachFinding1995}. We then compress this graph using sparse tables while preserving shortest-path costs.

\subsubsection{Fine-Grained Search Graph}\label{sec:fine-grained-search-graph}

The exchange graph $D$ is too large in two ways: an exchange $e\to f$ needs a matroid-level test, and the arcs $f\to e$ join every $f\in I$ to every $e\notin I$ with the same head.

We first replace the matroid exchanges by forest swaps.

\begin{definition}[Forest Swaps and Joining Elements]\leavevmode
	\begin{itemize}
		\item For $e\in E$ and $f\in I$, we call $e\to f$ a \emph{forest swap} in $F_i$ if $e\notin F_i$, the ends of $e$ are connected in $F_i$, and $f$ lies on the $F_i$-path between them.

		\item We say that $e$ \emph{reaches} $f$ by forest swaps if there are elements $e=e_0,e_1,\ldots,e_t=f$, with $t\ge0$, such that each $e_{j-1}\to e_j$ is a forest swap in some $F_i$. The forest may differ between steps.

		\item An element $e$ is \emph{joining} if its ends lie in different components of $F_i$ for some $i$ with $e\notin F_i$.
	\end{itemize}
\end{definition}

The next lemma relates forest-swap reachability to $\mathcal G^k$ exchanges and shows how to execute the corresponding paths. It specializes the standard matroid-union augmenting-path argument; see~\cite[Theorem~42.4]{schrijverCombinatorialOptimizationPolyhedra2004}.
\begin{lemma}[Forest-swap reachability and execution]
	\label{lem:forest-swap-reachability}
	Let $e\notin I$.
	\begin{enumerate}[label=(\alph*)]
		\item The element $e$ is free iff it reaches a joining element by forest swaps.
		\item If $e$ is nonfree, then the elements of $I$ reachable from $e$ are exactly the elements of $I$ in its fundamental circuit.
		\item If $e$ is nonfree, a path from $e$ to a reachable $f\in I$ with the fewest forest swaps yields a forest decomposition of $I-f+e$ in which each forest keeps its connected components.
	\end{enumerate}
\end{lemma}
We defer the proof to \Cref{app:forest-swap-reachability}.

\paragraph{Construction of the fine-grained graph $\FG$.}
Start with a vertex for every element of $E$ and a sink $z$. Each arc has a primary weight $w$ and a secondary length $q$. For every free element $e$, add an arc $e\to z$ with $(w,q)=(0,0)$.

We then add all forest swaps between elements of $E$. Formally, for each $e\in E$ and each forest $F_i$ in the decomposition with $e\notin F_i$, if the ends of $e$ are connected in $F_i$, we add an arc $e\to f$ with $(w,q)=(0,0)$ for every $f$ on the $F_i$-path between them. We call this operation \emph{scanning} $e$ into $F_i$. We iterate over all elements and forests and perform every applicable scan.

To represent the same-head exchanges of $D$, add a \emph{head vertex} $v^\star$ for every $v\in V$. For $f\in I$ and $e\notin I$, add $f\to\head(f)^\star:(-c(f),1), \head(e)^\star\to e:(c(e),0).$

For a shortest path, the goal is to minimize total weight first and total secondary length second.

\begin{lemma}[Path Correspondence between $D$ and $\FG$]\label{lem:fine-grained-projection}
	\leavevmode
	\begin{enumerate}[label=(\alph*)]
		\item Every simple $u^\star$-to-$z$ path $P_{\mathrm{fine}}$ in $\FG$ projects to an augmenting path $P$ from $u$ in $D$ with $w(P_{\mathrm{fine}})=\ell(P)$ and $q(P_{\mathrm{fine}})=q(P)$.
		\item Conversely, every augmenting path from $u$ with no free element before its last one lifts to a $u^\star$-to-$z$ walk in $\FG$ with the same cost and exchange count.
		\item If $I$ is extreme, then $\FG$ has no negative cycle.
	\end{enumerate}
\end{lemma}
\begin{proof}
	(a) To \emph{project} $P_{\mathrm{fine}}$ to $D$, delete the head vertices, $z$, and every element $e\in I$ that is not immediately followed by a head vertex. Let $P$ be the resulting sequence.

	Between consecutive head vertices, $P_{\mathrm{fine}}$ visits an element $e\notin I$ and follows forest swaps to an element $f\in I$. If $e$ is nonfree, $I-f+e$ is independent by \Cref{lem:forest-swap-reachability}(b); if $e$ is free, this holds for every $f\in I$. The segment therefore represents the exchange $e\to f$, with weight $c(e)-c(f)$ and secondary length one. The next element of $E\setminus I$ has the same head as $f$, giving a partition exchange of weight zero. The final segment enters a free element and then $z$, contributing the terminal cost. Thus $P$ is an augmenting path with the stated cost and exchange count; its elements are distinct because $P_{\mathrm{fine}}$ is simple.

	(b) To \emph{lift} an augmenting path $P$, replace each matroid exchange $e\to f$ by a sequence of forest swaps from $e$ to $f$. Such a sequence exists by \Cref{lem:forest-swap-reachability}(b), since every nonterminal element of $P\setminus I$ is nonfree. Replace each partition exchange $f\to e$ by $f\to\head(f)^\star\to e$, prepend $u^\star$, and append $z$. The resulting walk has the same cost and exchange count as $P$ by the accounting in~(a).

	(c) A cycle without head vertices has weight zero. A cycle with head vertices projects by the same argument to a closed walk of $D$ with the same weight, which is nonnegative when $I$ is extreme by \Cref{thm:weighted-matroid-intersection}.
\end{proof}

\subsubsection{Compressed Search Graph}
\label{sec:compressed-search-graph}\label{sec:tree-path-compression}

\paragraph{Tree-path compression.}

We use the tree-path compression of~\cite[Appendix~B]{saranurakIncrementalDirectedMinimum2026}, which we recall in our notation. The idea is to represent all forest swaps from an element to a tree path by at most four arcs into a shared auxiliary graph.

Root every component of each forest $F_i$. Write $p_i^d(x)$ for the ancestor $d$ edges above $x$ in $F_i$, and $f_i(x)$ for the edge joining a nonroot vertex $x$ to its parent. As in a sparse table, we create \emph{jump vertices} for power-of-two blocks of root paths.

For every $i,x,j$ with $j\ge0$ and $2^j\le\operatorname{depth}_i(x)$, create $x_{i,j}$ and add
\[
	\begin{aligned}
		x_{i,0} & \to f_i(x),          \\
		x_{i,j} & \to x_{i,j-1},\qquad
		x_{i,j}\to(p_i^{2^{j-1}}(x))_{i,j-1}\quad(j>0).
	\end{aligned}
\]
Give all these arcs weight zero. By induction on $j$, the elements of $I$ reachable from $x_{i,j}$ through jump vertices are exactly the $2^j$ edges from $x$ to $p_i^{2^j}(x)$. The auxiliary graph has $O(kn\log n)$ vertices and arcs.

\begin{figure}[tbp]
	\centering
	\resizebox{\linewidth}{!}{\begin{tikzpicture}[
				x=1cm,y=1cm,
				treevertex/.style={circle,fill=black,inner sep=1.35pt},
				treeedge/.style={line width=.8pt,gray!70},
				originaledge/.style={-{Stealth[length=1.8mm]},thick,dashed},
				edgevertex/.style={rectangle,draw,rounded corners=1pt,fill=gray!8,minimum width=6mm,minimum height=4.5mm,inner sep=1pt},
				virtual/.style={rectangle,draw=linkblue,rounded corners=1.5pt,fill=linkblue!6,minimum height=4.7mm,inner sep=1.5pt,align=center},
				tablevertex/.style={virtual,minimum width=8.5mm,minimum height=5.2mm,inner sep=1pt,font=\scriptsize},
				selected/.style={tablevertex,draw=linkblue!80!black,fill=linkblue!14,line width=.9pt},
				unused/.style={tablevertex,draw=gray!55,fill=gray!5,text=gray!65},
				zeroarc/.style={-{Stealth[length=1.5mm]},linkblue,line width=.55pt},
				unusedarc/.style={-{Stealth[length=1.4mm]},gray!42,line width=.45pt},
				leafarc/.style={-{Stealth[length=1.5mm]},citegreen,line width=.6pt},
				every node/.style={font=\small}
			]
			\node[treevertex,label=above:{$a=\mathrm{LCA}_i(u,v)$}] (a) at (1.9,2.55) {};
			\node[treevertex] (ul) at (.95,1.45) {};
			\node[treevertex] (vl) at (2.85,1.45) {};
			\node[treevertex,label=below:$u$] (u) at (.15,.35) {};
			\node[treevertex,label=below:$v$] (v) at (3.65,.35) {};
			\draw[treeedge] (a)--(ul)--(u);
			\draw[treeedge] (a)--(vl)--(v);
			\draw[originaledge] (u) to[bend right=28] node[below=2pt] {$e=(u,v)$} (v);
			\node[font=\scriptsize,rotate=49,fill=white,inner sep=.5pt] at (.78,1.38) {$u\leadsto a$};
			\node[font=\scriptsize,rotate=-49,fill=white,inner sep=.5pt] at (3.02,1.38) {$v\leadsto a$};
			\node[font=\footnotesize] at (1.9,-1.32) {(a) LCA decomposition};

			\begin{scope}[xshift=.35cm]
			\node[font=\scriptsize,text=gray!75!black,anchor=east] at (4.42,2.38) {$j=2$};
			\node[font=\scriptsize,text=gray!75!black,anchor=east] at (4.42,1.43) {$j=1$};
			\node[font=\scriptsize,text=gray!75!black,anchor=east] at (4.42,.48) {$j=0$};

			\node[selected] (x02) at (4.85,2.38) {$x_{0,2}$};
			\node[unused]   (x12) at (5.85,2.38) {$x_{1,2}$};
			\node[selected] (x22) at (6.85,2.38) {$x_{2,2}$};
			\node[unused]   (x32) at (7.85,2.38) {$x_{3,2}$};
			\node[unused]   (x42) at (8.85,2.38) {$x_{4,2}$};
			\node[unused]   (x52) at (9.85,2.38) {$x_{5,2}$};

			\node[tablevertex] (x01) at (4.85,1.43) {$x_{0,1}$};
			\node[unused]      (x11) at (5.85,1.43) {$x_{1,1}$};
			\node[tablevertex] (x21) at (6.85,1.43) {$x_{2,1}$};
			\node[unused]      (x31) at (7.85,1.43) {$x_{3,1}$};
			\node[tablevertex] (x41) at (8.85,1.43) {$x_{4,1}$};
			\node[unused]      (x51) at (9.85,1.43) {$x_{5,1}$};

			\foreach \r/\x in {0/4.85,1/5.85,2/6.85,3/7.85,4/8.85,5/9.85}{
				\node[tablevertex] (x\r0) at (\x,.48) {$x_{\r,0}$};
			}

			\node[edgevertex] (e) at (5.85,3.20) {$e$};
			\draw[zeroarc] (e)--node[pos=.48,left,yshift=3pt,inner sep=.5pt] {\scriptsize $0$} (x02);
			\draw[zeroarc] (e)--(x22);

			\draw[unusedarc] (x12)--(x11);
			\draw[unusedarc] (x12)--(x31);
			\draw[unusedarc] (x11)--(x10);
			\draw[unusedarc] (x11)--(x20);
			\draw[unusedarc] (x31)--(x30);
			\draw[unusedarc] (x31)--(x40);
			\draw[unusedarc] (x32)--(x31);
			\draw[unusedarc] (x32)--(x51);
			\draw[unusedarc] (x42)--(x41);
			\draw[unusedarc] (x42)--(10.85,1.43);
			\draw[unusedarc] (x52)--(x51);
			\draw[unusedarc] (x52)--(11.85,1.43);
			\draw[unusedarc] (x51)--(x50);
			\draw[unusedarc] (x51)--(10.85,.48);

			\draw[zeroarc] (x02)--(x01);
			\draw[zeroarc] (x02)--(x21);
			\draw[zeroarc] (x22)--(x21);
			\draw[zeroarc] (x22)--(x41);
			\draw[zeroarc] (x01)--(x00);
			\draw[zeroarc] (x01)--(x10);
			\draw[zeroarc] (x21)--(x20);
			\draw[zeroarc] (x21)--(x30);
			\draw[zeroarc] (x41)--(x40);
			\draw[zeroarc] (x41)--(x50);

			\foreach \j/\x in {0/4.35,1/5.35,2/6.35,3/7.35,4/8.35,5/9.35,6/10.35}{
				\node[treevertex] (p\j) at (\x,-.38) {};
			}
			\draw[treeedge] (p0)--(p1)--(p2)--(p3)--(p4)--(p5)--(p6);
			\foreach \h/\x in {1/4.85,2/5.85,3/6.85,4/7.85,5/8.85,6/9.85}{
				\node[edgevertex] (f\h) at (\x,-.38) {$f_{\h}$};
			}
			\foreach \r/\h in {0/1,1/2,2/3,3/4,4/5,5/6}{
				\draw[leafarc] (x\r0)--(f\h);
			}
			\node[font=\scriptsize,citegreen,fill=white,inner sep=.5pt] at (4.62,.03) {$0$};
			\node[font=\scriptsize,anchor=north] at (4.35,-.56) {$u=x_0$};
			\node[font=\scriptsize,anchor=north] at (10.35,-.56) {$a=x_6$};
			\draw[gray!65,dashed,-{Stealth[length=1.5mm]}] (p6)--(11.15,-.38);
			\node[font=\tiny,text=gray!75!black,anchor=south] at (10.75,-.33) {rootward};
			\node[font=\scriptsize,text=gray!75!black,anchor=west] at (11.22,-.38) {$\mathrm{root}_i$};
			\node[font=\footnotesize,align=center] at (8.3,-1.32)
			{(b) Auxiliary connections and vertices reachable from $e$\\[-1pt]
				when the $u$--$a$ path has six edges};
			\end{scope}
		\end{tikzpicture}}
	\caption{Tree-path compression, adapted from~\cite[Figure~2]{saranurakIncrementalDirectedMinimum2026}. (a) Split the scanned path in $F_i$ at its lowest common ancestor $a$. (b) For a six-edge $u$--$a$ branch, write $x_r=p_i^r(u)$, $f_{r+1}=f_i(x_r)$, and $x_{r,j}=(p_i^r(u))_{i,j}$. The two entry blocks $x_{0,2}$ and $x_{2,2}$ cover $f_1,\ldots,f_4$ and $f_3,\ldots,f_6$, respectively, and share $x_{2,1}$. Blue entries and arcs show the reachable jump structure; green arcs lead to forest elements. Gray entries are not reached from these two blocks. All auxiliary arcs have weight zero.}
	\label{fig:tree-path-compression}
\end{figure}
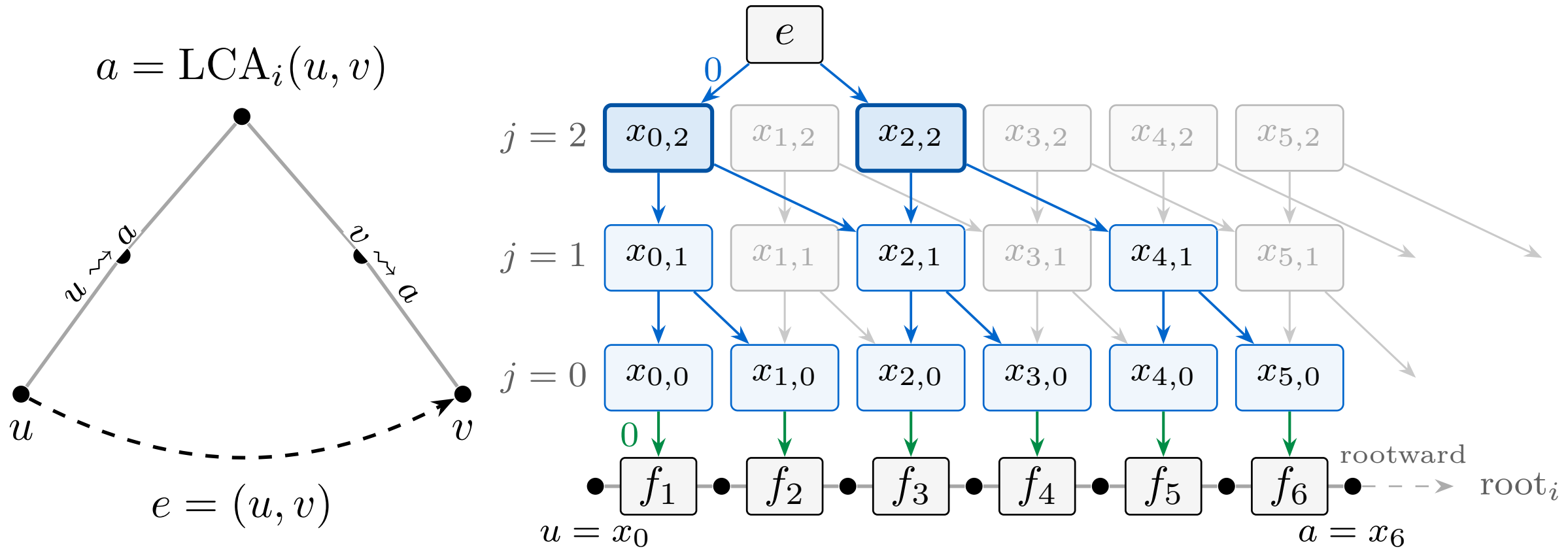

Now let $e=(u,v)$ scan its path in $F_i$, and let $a$ be the lowest common ancestor of $u$ and $v$. For each nonempty branch from $x\in\{u,v\}$ to $a$, let $d$ be its length and put $j=\lfloor\log_2d\rfloor$. Add weight-zero arcs from $e$ to the two jump vertices
\begin{equation}\label{eq:two-block-cover}
	x_{i,j},\qquad (p_i^{d-2^j}(x))_{i,j},
\end{equation}
omitting one if they coincide. Since $2^j\le d<2^{j+1}$, the two length-$2^j$ blocks cover the entire branch. Thus at most four arcs represent all forest swaps from $e$ into $F_i$; see \Cref{fig:tree-path-compression}. A route from $e$ to an element $f\in I$, with only jump vertices in between, exists exactly when $f$ lies on the scanned path, and has weight zero. Costs and exchange counts remain on the arcs incident to head vertices.

We build the arcs between jump vertices and preprocess each forest for $O(1)$-time LCA queries in $O(kn\log n)$ total time, using Euler tours and sparse tables. For each scanned tree path, its two-block cover in~\eqref{eq:two-block-cover} requires only a constant number of ancestor queries. We answer all ancestor queries offline in one depth-first traversal of each forest: maintain the current root path in an array indexed by depth, and answer each query at $x$ when the traversal visits $x$. Thus, if the elements scan $s$ tree paths in total, the scan graph has $O(s+kn\log n)$ arcs and can be built in the same time.

\paragraph{Construction of the compressed graph.}

To build $\CG$, replace the forest swaps of $\FG$ by the jump construction above.

We then suppress the vertex for every element $e\notin I$: we combine its only incoming arc, from $\head(e)^\star$, with each outgoing arc and label each resulting arc by $e$, so that we can recover $e$ when reconstructing an augmenting path. The remaining vertices are those for elements of $I$, the head vertices, the jump vertices, and $z$. This reduces the number of vertices from $O(m+kn\log n)$ to $O(kn\log n)$, avoiding an additional logarithmic factor on the $m$ term in Dijkstra's running time.

For a scanning element $e$, let $J(e)$ be the (at most $4k$) jump vertices covering its scanned tree paths as in~\eqref{eq:two-block-cover}. The resulting arcs are listed below; we keep parallel arcs as they can have different labels.

\begin{center}
	\small
	\begin{tabular}{@{}llccc@{}}
		\toprule
		type          & arc                                             & $w$     & $q$ & label \\
		\midrule
		Head exchange & $f\to\head(f)^\star$, $f\in I$                  & $-c(f)$ & $1$ & ---   \\
		              & $\head(e)^\star\to x$, $e\notin I$, $x\in J(e)$ & $c(e)$  & $0$ & $e$   \\
		\addlinespace[4pt]
		Forest scans  & $f\to x$, $f\in I$, $x\in J(f)$                 & $0$     & $0$ & ---   \\
		              & jump vertex $\to$ child                         & $0$     & $0$ & ---   \\
		              & $x_{i,0}\to f_i(x)$                             & $0$     & $0$ & ---   \\
		\addlinespace[4pt]
		To the sink   & $\head(e)^\star\to z$, $e$ free                 & $c(e)$  & $0$ & $e$   \\
		\bottomrule
	\end{tabular}
\end{center}

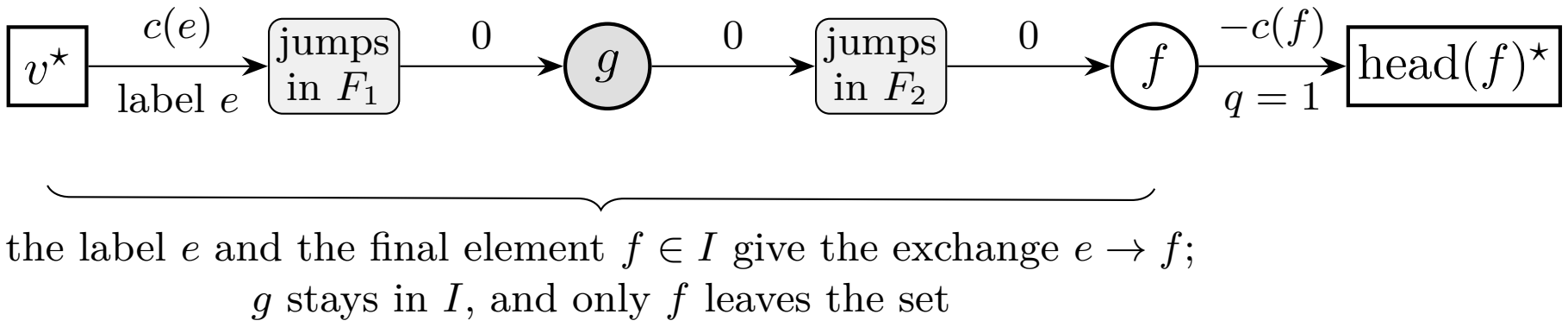
\begin{figure}[htb]
	\centering
	\begin{tikzpicture}[>={Stealth[length=2mm]},
			el/.style={circle,draw,thick,minimum size=6.5mm,inner sep=0pt},
			headvertex/.style={rectangle,draw,thick,minimum size=6mm,inner sep=2pt},
			box/.style={rectangle,rounded corners=3pt,draw,fill=gray!12,minimum height=7mm,align=center,font=\footnotesize,inner sep=2pt},
			lab/.style={font=\footnotesize,above}]
		\node[headvertex] (hv) at (0,0) {$v^\star$};
		\node[box] (s1) at (2.2,0) {jumps\\[-1pt] in $F_1$};
		\node[el,fill=gray!25] (g) at (4.3,0) {$g$};
		\node[box] (s2) at (6.4,0) {jumps\\[-1pt] in $F_2$};
		\node[el] (f) at (8.5,0) {$f$};
		\node[headvertex] (hf) at (10.8,0) {$\head(f)^\star$};
		\draw[->] (hv) -- node[lab]{$c(e)$} node[below,font=\footnotesize]{label $e$} (s1);
		\draw[->] (s1) -- node[lab]{$0$} (g);
		\draw[->] (g) -- node[lab]{$0$} (s2);
		\draw[->] (s2) -- node[lab]{$0$} (f);
		\draw[->] (f) -- node[lab]{$-c(f)$} node[below,font=\footnotesize]{$q=1$} (hf);
		\draw[decorate,decoration={brace,mirror,amplitude=5pt}] ($(hv.south)+(0,-0.6)$) -- ($(f.south)+(0,-0.6)$)
		node[midway,below=6pt,font=\footnotesize,align=center]{the label $e$ and the final element $f\in I$ give the exchange $e\to f$;\\ $g$ stays in $I$, and only $f$ leaves the set};
	\end{tikzpicture}
	\caption{A path segment between two head vertices in the compressed graph. The label $e\notin I$ records an element entering $v$ and scanning $F_1$. The element $g\in F_1$ scans $F_2$. Only the arcs incident to head vertices carry costs, and only the arc into a head vertex counts as an exchange for the tie-breaker $q$.}
	\label{fig:compressed-path}
\end{figure}

\begin{lemma}[Search in the compressed graph]\label{lem:compressed-graph-search}
	The statements of \Cref{lem:fine-grained-projection} hold for $\CG$ as well. If $I$ is extreme, the distance from each $v^\star$ to $z$ equals the minimum cost of an augmenting path from $v$ in $D$.
\end{lemma}
\begin{proof}
	To \emph{project} a simple $u^\star$-to-$z$ path in $\CG$ to $D$, first insert each arc label $e\notin I$ just after its head vertex. Then delete the head vertices, $z$, the jump vertices, and every element $f\in I$ not immediately followed by a head vertex. See \Cref{fig:compressed-path} for an example.

	Each jump route represents exactly one forest swap, so the same argument as in \Cref{lem:fine-grained-projection}(a) shows that the projection preserves the cost and exchange count. Its elements of $I$ are distinct because the search path is simple. Its labels from $E\setminus I$ are also distinct: every arc labeled by $e$ leaves the same vertex $\head(e)^\star$, which the path visits at most once. Hence the projection is an augmenting path.

	Conversely, \emph{lift} an augmenting path with no nonterminal free element to a walk in $\FG$ using \Cref{lem:fine-grained-projection}(b). Replace each forest swap by a jump route and suppress each vertex $e\notin I$, keeping $e$ as a label. This gives a walk in $\CG$ with the same cost and exchange count.

	Expanding a cycle in $\CG$ gives a closed walk in $\FG$ with the same weight. Thus $\CG$ has no negative cycle when $I$ is extreme, by \Cref{lem:fine-grained-projection}(c). Finally, a shortest augmenting path has no nonterminal free element by \Cref{lem:no-nonterminal-free-elements}, so it lifts to a walk in $\CG$. Conversely, a shortest simple search path projects to an augmenting path of the same cost. These two directions give the distance equality.
\end{proof}

\paragraph{Restricting scans to one forest.}
So far, an element may scan up to $k$ forests. Following the cyclic-scanning idea of Gabow and Westermann~\cite[Lemma~2.2]{gabowForestsFramesGames1992}, we restrict each element to scan only one forest. We now let every element of $E\setminus I$ scan only $F_1$, and every element of $F_i$ scan only $F_{i+1}$, with indices modulo $k$; all other arcs and weights stay the same. For $k=1$, elements of $I$ do not scan. Thus $|J(e)|\le 4$, and these jump vertices cover at most one tree path.

These cyclic scans preserve the elements of $I$ reachable from each nonfree element $e\notin I$; we defer the proof to \Cref{app:cyclic-scans}. This is the reachability property used above, so \Cref{lem:compressed-graph-search} remains valid. From now on, $\CG$ denotes this restricted graph.

Each element now scans at most one tree path, and the jump graphs of the $k$ forests have $O(kn\log n)$ total size. We can therefore build $\CG$ in $O(m+kn\log n)$ time, with
\begin{equation}\label{eq:compressed-graph-size}
	N=O(kn\log n)\ \text{vertices},\qquad M=O(m+kn\log n)\ \text{arcs}.
\end{equation}

\paragraph{Shortest paths and path recovery.}
Suppose that $I$ is extreme. Compute exact distances $\hat d(x)$ to $z$ by one reverse SSSP computation on $\CG$. We then keep only arcs satisfying $\hat d(x)=w(x,y)+\hat d(y)$, and run a reverse $0$--$1$ BFS from $z$ with lengths $q$. This gives a shortest-path tree that minimizes the number of exchanges among minimum-cost paths. For each chosen $u_T$ with finite distance, project its tree path to $D$ to obtain the proposal $P_T$, by \Cref{lem:compressed-graph-search}.

The work after SSSP takes $O(M+N)$ total time. Filtering and BFS are linear. To bound the total length of the tree paths, note that their projections have no nonterminal free element. Thus the scans of the path from $u_T$ stay inside $T$, and each visited jump vertex represents a nonempty tree-path segment inside $T$. Paths from distinct blocks are therefore vertex-disjoint except at $z$, so projecting takes $O(N)$ total time as well.

\subsection{Finding a Maximum Compatible Set of Proposals}\label{sec:compatible-set-selection}
Let $H=\Ix\dotcup A$ be the arc set of the contracted graph in \Cref{sec:contracted-compatibility}, and let $t=|\cT(I)|$. We call the arcs in $\Ix$  \emph{mandatory}. The algorithm of \Cref{lem:ak-forest-packing} gives a basis $B$ of $H$, possibly missing some arcs of $\Ix$. We repair $B$ by restoring these arcs one at a time, each time removing an arc outside $\Ix$ and updating the forest decomposition by forest swaps. In this subsection all forests and tight blocks are in the contracted graph.

\begin{algorithm}[H]
	\caption{$\textsc{Extend}(H)$: largest independent $K$ with $\Ix\subseteq K\subseteq H$}\label{alg:mandatory-extension}
	\begin{algorithmic}[1]
		\Require the arc set $H=\Ix\dotcup A$ of the contracted graph on $t$ vertices, with indegree at most $k$, where $\Ix$ is independent in $\mathcal G^k$
		\Ensure a largest independent $K$ with $\Ix\subseteq K\subseteq \Ix\cup A$, and a forest decomposition of $K$
		\State compute a basis $B$ of $H$ with forest decomposition $F_1,\ldots,F_k$ using \Cref{lem:ak-forest-packing}
		\State compute $\cT(B)$
		\For{each block $T\in\cT(B)$}
		\While{some $e\in \Ix[T]$ is not in $B$}\label{line:mandatory-extension-while}
		\State build a graph representing all forest swaps among $\{e\}\cup B[T]$, using forest scans and the tree-path compression of \Cref{sec:tree-path-compression}
		\State find a path $e=e_0\to\cdots\to e_s$ of forest swaps with $e_s\notin \Ix$ and $s$ minimum \Comment{$0$--$1$ BFS}
		\State for all $1\le j\le s$ at once: remove $e_j$ from its old forest $F_{i_j}$ and insert $e_{j-1}$ into $F_{i_j}$
		\State $B\gets B-e_s+e_0$
		\EndWhile
		\EndFor
		\State \Return $K\gets B$ with its current forest decomposition
	\end{algorithmic}
\end{algorithm}

\begin{lemma}[Restoring a mandatory arc]\label{lem:mandatory-insertion}
	Let $T$ be a $B$-tight block and let $e\in\Ix[T]\setminus B$. Given the spanning trees $B[T]=F_1[T]\dotcup\cdots\dotcup F_k[T]$, we can find an arc $f\in B[T]\setminus\Ix$ and a decomposition of $B[T]-f+e$ into $k$ spanning trees of $T$ in $O(k^2|T|+k|T|\log|T|)$ time.
\end{lemma}
\begin{proof}
	The fundamental circuit of $e$ with respect to $B[T]$ is not contained in the independent set $\Ix$ (as $e \in \Ix$), so it contains an element of $B[T]\setminus\Ix$. By \Cref{lem:forest-swap-reachability}(b), there is a path of forest swaps from $e$ to $B[T]\setminus\Ix$. We find such a path $e=e_0\to\cdots\to e_s=f$ with the fewest swaps. By \Cref{lem:forest-swap-reachability}(c), executing these forest swaps gives a decomposition of $B[T]-f+e$ with the same forest components. Thus each forest remains a spanning tree of $T$, and all elements of $B[T]\cap\Ix$ are preserved.

	For the running time, we use the tree-path compression of \Cref{sec:tree-path-compression} without the cyclic-scanning restriction: $e$ and every element of $B[T]$ scan their paths in all trees $F_i[T]$ that do not contain them. The resulting graph has $O(k^2|T|+k|T|\log|T|)$ arcs. A $0$--$1$ BFS, with length one on arcs leaving element vertices, finds a path with the fewest swaps in time linear in the size of this graph.
\end{proof}

\begin{lemma}\label{lem:mandatory-extension}
	\Cref{alg:mandatory-extension} is correct and runs in $O(k^2(|H|+t)\log(kt)+k^3t)$ time.
\end{lemma}
\begin{proof}
	\emph{Correctness.}
	Since every vertex of the contracted graph has indegree at most $k$, \Cref{lem:ak-forest-packing} guarantees that the initial forest packing $B$ is a basis of $H$. An element of $H\setminus B$ that crosses $\cT(B)$ would be free for $B$ (\Cref{lem:free-elements}), so all missing mandatory arcs lie inside blocks.

	Each iteration of the inner loop repairs one missing mandatory arc using \Cref{lem:mandatory-insertion}: it adds that arc, removes an arc outside $\Ix$, and keeps every forest restricted to $T$ a spanning tree. Therefore, after the while loop (\cref{line:mandatory-extension-while}), $B[T]$ contains $\Ix[T]$.

	Each repair replaces, for each $i$, the spanning tree $F_i[T]$ by another spanning tree on $T$, leaving all other arcs unchanged. Thus every $F_i$ remains a forest. The final set contains $\Ix$ and has the same cardinality as the initial basis, so it is a largest independent set containing $\Ix$.

	\emph{Running time.}
	The initial basis computation takes $O(k^2(|H|+t)\log(kt))$ time by \Cref{lem:ak-forest-packing}. Each block $T$ is missing at most $k$ mandatory arcs: the heads in $T$ receive at most $k|T|$ arcs of $H$, whereas $|B[T]|=k(|T|-1)$. Thus each block requires at most $k$ repairs, taking
	\[
		O\Bigl(\sum_T k(k^2|T|+k|T|\log|T|)\Bigr)
		=O(k^3t+k^2t\log t)
	\]
	time in total.
\end{proof}

We implement the extension step of \Cref{alg:batching-framework} by computing $K\gets\textsc{Extend}(H)$ and taking $A'=K\setminus\Ix$. Since $|H|\le k(t-1)$ by \Cref{lem:contracted-indegrees} and $t\le n$, the extension takes $O(k^3n\log(kn))$ time per phase.

\subsection{Proof of the Relaxation Theorem}\label{sec:relaxation-proof}

We return to the matroid-and-partition setting of \Cref{sec:relax-and-augment} and use one more fact from~\cite{schrijverCombinatorialOptimizationPolyhedra2004}. Our exchange graph is the reverse of the one there, and the weight $w(C)$ of a cycle, i.e., the total weight of its arcs, is its length there for the weights $-c$.

\begin{lemma}[{\cite[Lemma~41.5$\alpha$]{schrijverCombinatorialOptimizationPolyhedra2004}}]
	\label{lem:cycle-shortening}
	Let $J$ be common independent in $M_1$ and $M_2$, let $C$ be a cycle in $D_{M_1,M_2}(J)$, and let $t\in V(C)$. If $J\oplus V(C)$ is not common independent, then $D_{M_1,M_2}(J)$ has a cycle $C'$ with $V(C')\subsetneq V(C)$ such that $w(C')<0$, or $w(C')\le w(C)$ and $t\in V(C')$.
\end{lemma}

\begin{proof}[Proof of \Cref{thm:relaxation}]
	We first verify that an augmenting path from class $u$ exists whenever $\OPT(\vb+\one_u)<+\infty$. In this case, $I$ is not a maximum-cardinality common independent set of $M$ and $N_{\vb+\one_u}$. By \Cref{thm:weighted-matroid-intersection}, there is a path from $E_u\setminus I$ to $X_1(I)$ in $D_{M,N_{\vb+\one_u}}(I)$. Every arc of this graph that is absent from $D(I)$ enters $E_u\setminus I$. The suffix starting at the last vertex in $E_u\setminus I$ therefore lies in $D(I)$ and is an augmenting path from class $u$.

	For every such $u$, define the minimum augmenting-path cost
	\[
		\widehat{\Delta}_u=\min\{\ell(P):P\text{ is an augmenting path from class }u\}.
	\]
	We will prove that $\widehat{\Delta}_u=\Delta_u$.

	We adapt the virtual-element construction in the proof of~\cite[Theorem~41.6]{schrijverCombinatorialOptimizationPolyhedra2004}, adding one virtual element for each relaxed class.
	Fix $U\subseteq V$ such that $\OPT(\vb+\one_u)<+\infty$ for every $u\in U$. For each $u\in U$, let $\theta_u$ be a new virtual element of cost $\widehat{\Delta}_u$, and let $\Theta=\{\theta_u:u\in U\}$. We extend $M$ and $N_{\vb}$ to matroids $M'$ and $N'$ on $E\cup\Theta$ as follows:
	\begin{itemize}
		\item A set $X$ is independent in $M'$ iff $X\setminus\Theta$ is independent in $M$.
		\item $N'$ is the partition matroid obtained by adding $\theta_u$ to class $u$ and increasing its capacity by one, for every $u\in U$.
	\end{itemize}
	Then $I+\Theta$ is common independent in $M'$ and $N'$, and every class is full. Let $D'=D_{M',N'}(I+\Theta)$. \Cref{fig:virtual-elements} illustrates this construction.

	\begin{figure}[H]
		\centering
		\resizebox{\linewidth}{!}{\begingroup
\colorlet{wapM}{linkblue}
\colorlet{wapP}{urlorange}
\colorlet{wapVirtual}{citegreen}
\definecolor{wapInk}{HTML}{303842}
\definecolor{wapHighlight}{HTML}{FFE45C}
\begin{tikzpicture}[
    x=1cm,y=1cm,font=\small,
    >={Stealth[length=1.8mm,width=1.25mm]},
    ground/.style={draw=wapInk!45,line width=.6pt,fill=black!1},
    selectedset/.style={draw=wapInk!60,line width=.7pt,fill=black!6},
    freeset/.style={draw=wapM,line width=.8pt,fill=wapM!7},
    classset/.style={draw=black,line width=.8pt,dashed,fill=none},
    outside/.style={circle,draw=wapInk,fill=white,line width=.65pt,
                    inner sep=0pt,minimum size=4.4pt},
    selected/.style={circle,draw=wapInk,fill=wapInk,
                     inner sep=0pt,minimum size=4.4pt},
    virtual/.style={circle,draw=wapVirtual,fill=wapVirtual,
                    inner sep=0pt,minimum size=4.6pt},
    mpath/.style={->,draw=wapM,line width=1.25pt},
    mfan/.style={->,draw=wapM!43,line width=.65pt},
    ppath/.style={->,draw=wapP,line width=1.25pt},
    pathhighlight/.style={draw=wapHighlight,opacity=.75,line width=8.5pt,
                          line cap=round,line join=round},
    vertexlabel/.style={text=wapInk,inner sep=1.5pt},
    heading/.style={font=\small\bfseries,text=wapInk},
    legend/.style={font=\footnotesize,text=wapInk,anchor=west}
]
\foreach \shift/\panel in {0/0,8.4/1} {
  \begin{scope}[xshift=\shift cm]
    \path[ground] (0,-.35) ellipse[x radius=3.85,y radius=2.90];
    \ifnum\panel=0
      \path[selectedset] (0,-1.05) ellipse[x radius=2.28,y radius=1.00];
    \else
      \path[fill=black!6]
        (2.28,-1.05)
        .. controls (2.28,-1.54) and (1.61,-1.88) .. (.95,-1.97)
        .. controls (.98,-2.54) and (.58,-2.86) .. (0,-2.86)
        .. controls (-.58,-2.86) and (-.98,-2.54) .. (-.95,-1.97)
        .. controls (-1.61,-1.88) and (-2.28,-1.54) .. (-2.28,-1.05)
        .. controls (-2.28,-.50) and (-1.26,-.05) .. (0,-.05)
        .. controls (1.26,-.05) and (2.28,-.50) .. (2.28,-1.05) -- cycle;
      \path[fill=wapVirtual!10]
        (-.95,-1.97) .. controls (-.50,-2.07) and (.50,-2.07) .. (.95,-1.97)
        .. controls (.98,-2.54) and (.58,-2.86) .. (0,-2.86)
        .. controls (-.58,-2.86) and (-.98,-2.54) .. (-.95,-1.97) -- cycle;
      \draw[draw=wapInk!60,line width=.7pt]
        (2.28,-1.05)
        .. controls (2.28,-1.54) and (1.61,-1.88) .. (.95,-1.97)
        .. controls (.98,-2.54) and (.58,-2.86) .. (0,-2.86)
        .. controls (-.58,-2.86) and (-.98,-2.54) .. (-.95,-1.97)
        .. controls (-1.61,-1.88) and (-2.28,-1.54) .. (-2.28,-1.05)
        .. controls (-2.28,-.50) and (-1.26,-.05) .. (0,-.05)
        .. controls (1.26,-.05) and (2.28,-.50) .. (2.28,-1.05);
      \draw[draw=wapVirtual!55,densely dashed,line width=.6pt]
        (-.95,-1.97) .. controls (-.50,-2.07) and (.50,-2.07) .. (.95,-1.97);
      \node[text=wapVirtual] at (-.08,-2.25) {$\Theta$};
      \node[virtual] (theta) at (.60,-2.52) {};
      \node[virtual] (otherTheta) at (-.60,-2.52) {};
      \node[text=wapVirtual,font=\scriptsize] at (-.10,-2.50) {$\cdots$};
    \fi

    \path[freeset] (-2.20,1.00) ellipse[x radius=.90,y radius=.68];
    \path[classset] (2.20,1.00) ellipse[x radius=.78,y radius=.68];
    \node[text=black] at (2.20,1.43) {$E_u\setminus I$};
    \node[text=wapInk] at (0,-1.82) {$I$};

    \coordinate (e0pos) at (2.20,.84);
    \coordinate (e1pos) at (0,1.02);
    \coordinate (e2pos) at (-1.88,.99);
    \coordinate (f0pos) at (1.05,-.40);
    \coordinate (f1pos) at (-1.05,-.40);
    \draw[pathhighlight] (e0pos) -- (f0pos) -- (e1pos) -- (f1pos) -- (e2pos);
    \node[outside] (e0) at (e0pos) {};
    \node[outside] (e1) at (e1pos) {};
    \node[outside] (e2) at (e2pos) {};
    \node[selected] (f0) at (f0pos) {};
    \node[selected] (f1) at (f1pos) {};
    \node[selected] (g0) at (-.80,-1.15) {};
    \node[selected] (g1) at (.85,-1.40) {};

    \draw[mfan] (e2)
      .. controls (-1.90,.15) and (-1.62,-.55) .. (f1);
    \draw[mfan] (e2)
      .. controls (-2.20,-.40) and (.10,-1.55) .. (f0);
    \draw[mfan] (e2)
      .. controls (-2.70,-.55) and (-1.70,-1.20) .. (g0);
    \draw[mfan] (e2)
      .. controls (-3.05,-1.00) and (-1.05,-2.00) .. (g1);

    \ifnum\panel=1
      \draw[mfan] (e2)
        .. controls (-3.20,-.90) and (-1.85,-2.40) .. (otherTheta);
      \draw[mpath] (e2)
        .. controls (-2.18,1.18) and (-3.20,1.09) .. (-3.22,.30)
        .. controls (-3.40,-1.65) and (-1.70,-3.26) .. (theta);
      \draw[ppath] (theta)
        .. controls (3.14,-2.65) and (3.27,-.22) .. (e0);
      \node[vertexlabel,text=wapVirtual,above=3pt] at (theta) {$\theta_u$};
    \fi

    \draw[mpath] (e0) -- (f0);
    \draw[ppath] (f0) -- (e1);
    \draw[mpath] (e1) -- (f1);
    \draw[ppath] (f1) -- (e2);

    \node[vertexlabel,right=3pt] at (e0) {$e_0$};
    \node[vertexlabel,above=3pt] at (e1) {$e_1$};
    \node[vertexlabel,right=3pt] at (e2) {$e_2$};
    \node[vertexlabel,right=3pt] at (f0) {$f_0$};
    \node[vertexlabel,right=3pt] at (f1) {$f_1$};

    \ifnum\panel=0
      \node[heading] at (0,3.28) {(a) Original exchange graph};
      \node at (0,2.89) {$D_{M,N_{\bm b}}(I)$};
      \node[text=wapM] at (-2.20,1.40) {$X_1(I)$};
      \node[text=wapInk!75,font=\footnotesize] at (0,-3.48) {$E$};
    \else
      \node[heading] at (0,3.28) {(b) Add virtual elements};
      \node at (0,2.89) {$D_{M',N'}(I+\Theta)$};
      \node[text=wapM] at (-2.20,1.40) {$X_1(I+\Theta)$};
      \node[text=wapInk!75,font=\footnotesize] at (0,-3.48) {$E\cup\Theta$};
    \fi
  \end{scope}
}
\node[text=wapInk!65] at (4.2,.15) {$\Longrightarrow$};
\draw[mpath] (-2.80,-4.00) -- (-2.05,-4.00);
\node[legend] at (-1.88,-4.00)
  {$M$- or $M'$-exchange: weight $c(e)-c(f)$};
\draw[ppath] (5.15,-4.00) -- (5.90,-4.00);
\node[legend] at (6.07,-4.00)
  {$N_{\bm b}$- or $N'$-exchange: weight $0$};

\node[font=\footnotesize,text=wapInk] (pathlegend) at (4.65,-4.52)
  {Augmenting path $P_u:\ e_0\to f_0\to e_1\to f_1\to e_2$};
\draw[pathhighlight] ([xshift=-1cm]pathlegend.west) -- ([xshift=-.25cm]pathlegend.west);

\end{tikzpicture}
\endgroup}
		\caption{The virtual-element construction. The yellow band marks an augmenting path $P_u$ from class $u$; the two outer arcs extend $P_u$ to a cycle through $\theta_u$. Every element of $X_1(I)$ has an $M$-exchange arc to every element of $I$; in the right panel, every element of $X_1(I+\Theta)$ has an $M'$-exchange arc to every element of $I+\Theta$.}
		\label{fig:virtual-elements}
	\end{figure}
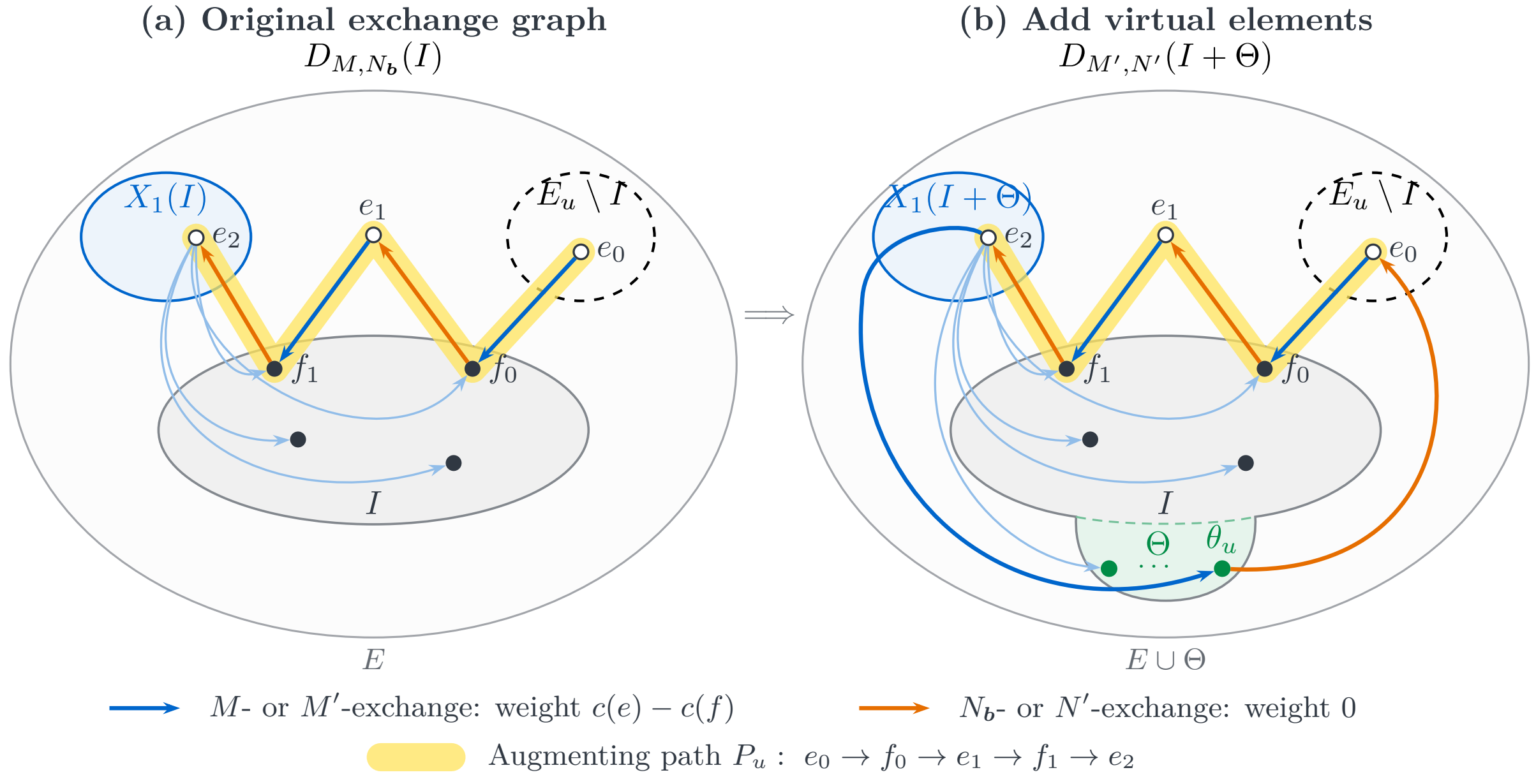

	\begin{claim*}
		The set $I+\Theta$ is extreme for $M'$ and $N'$, and
		\begin{equation}
			\OPT(\vb+\one_U)\ge\OPT(\vb)+\sum_{u\in U}\widehat{\Delta}_u.
			\label{eq:relaxation-lower-bound}
		\end{equation}
	\end{claim*}
	\begin{proof}[Proof of the claim]
		By \Cref{thm:weighted-matroid-intersection}, it suffices for extremality to show that $D'$ has no negative cycle. Compared with $D(I)$, the graph $D'$ has the new vertices $\theta_u$ and the following additional arcs:
		\[
			\theta_u\to e\text{ of weight }0
			\quad(e\in E_u\setminus I),\qquad
			e\to\theta_u\text{ of weight }c(e)-\widehat{\Delta}_u
			\quad(e\in X_1(I)).
		\]
		Indeed, $(I+\Theta)-\theta_u+e$ is independent in $M'$ iff $e\in X_1(I)$, and in $N'$ iff $e\in E_u$.

		Since $D(I)$ has no negative cycle, any negative cycle in $D'$ must pass through $\Theta$. Split such a cycle at its virtual elements into pieces $\theta_{u}\to P_u\to\theta_{v},$ where $P_u$ is an augmenting path from class $u$ in $D(I)$. This piece has weight $\ell(P_u)-\widehat{\Delta}_{v}$. Each piece ends at the virtual element where the next piece begins (cyclically), so the total weight is
		\[
			\sum_{\text{pieces}}\bigl(\ell(P_u)-\widehat{\Delta}_v\bigr)
			=\sum_{\text{pieces}}\bigl(\ell(P_u)-\widehat{\Delta}_u\bigr)\ge0.
		\]
		Thus the cycle has nonnegative weight, and $I+\Theta$ is extreme.

		Every independent set $X$ of $M$ with class counts $\vb+\one_U$ is also common independent in $M'$ and $N'$, with $|X|=|I+\Theta|$. Hence
		\[
			c(X)\ge c(I+\Theta)=\OPT(\vb)+\sum_{u\in U}\widehat{\Delta}_u,
		\]
		which proves the lower bound.
	\end{proof}

	\emph{Part (a).} Let $P=(e_0,f_0,\ldots,f_{q-1},e_q)$ be any shortest augmenting path from class $u$, so $\ell(P)=\widehat{\Delta}_u$. Take $U=\{u\}$ in the construction above. By the claim, $I+\theta_u$ is extreme, and $D'$ has no negative cycle.

	We next show that $I\oplus P$ is independent. Extend $P$ by the arcs $\theta_u\to e_0$ and $e_q\to\theta_u$ to form a cycle $C$. Then $w(C)=\ell(P)-\widehat{\Delta}_u=0$ and $(I+\theta_u)\oplus V(C)=I\oplus P$. If this set were not common independent in $M'$ and $N'$, \Cref{lem:cycle-shortening} with $t=\theta_u$ would give a cycle $C'$ through $\theta_u$ with $V(C')\subsetneq V(C)$ and $w(C')=0$, since $D'$ has no negative cycle. Removing $\theta_u$ from $C'$ would leave an augmenting path from class $u$ of cost $\ell(P)$ with fewer exchanges, contradicting the tie-breaking rule for $P$. Thus $I\oplus P$ is independent in $M$ with class counts $\vb+\one_u$.

	Its cost is $c(I)+\ell(P)=\OPT(\vb)+\widehat{\Delta}_u$, matching the lower bound~\eqref{eq:relaxation-lower-bound} for $U=\{u\}$. Hence $\ell(P)=\widehat{\Delta}_u=\Delta_u$.

	\emph{Part (b).} If $\OPT(\vb+\one_U)=+\infty$, the inequality is immediate. Otherwise, for each $u\in U$, relaxing $u$ alone is feasible, i.e., $\OPT(\vb+\one_u)<+\infty$. By part~(a), $\widehat{\Delta}_u=\Delta_u$ for every $u\in U$, so \eqref{eq:relaxation-lower-bound} gives the desired inequality.
\end{proof}

\section{Strongly Polynomial Algorithm via Vertex Prices}\label{sec:vertex-prices}

We now complete the algorithm and prove \Cref{thm:main}. We maintain vertex prices that certify optimality of the current set. The vertex prices can be viewed as dual variables for the current indegree equalities. We first show how to update them so that any compatible batch preserves the certificate, and then extend them to a feasible potential on the compressed graph. This lets us find the required paths using Dijkstra's algorithm within the batching framework of \Cref{sec:arborescence-packing}.

This shares the spirit of the classical weight-splitting approach to weighted matroid intersection~\cite{frankWeightedMatroidIntersection1981}; see also~\cite{schrijverCombinatorialOptimizationPolyhedra2004,iriOptimalIndependentAssignment1976}. For general matroid intersection, the split weights are attached to elements. In bipartite matching, where both matroids are partition matroids, they can be attached to the vertices, that is, to the classes of the partitions, on both sides. In our setting, one side is a partition matroid, so we naturally get vertex prices on that side. We believe the invariant and update below also apply to an arbitrary matroid intersected with a partition matroid.

\subsection{Price Invariant}\label{sec:price-invariant}

For vertex prices $h:V\to\mathbb R$, the \emph{adjusted cost} of an arc is $a(e)=c(e)-h(\head(e))$, and $a(X)=\sum_{e\in X}a(e)$. The algorithm keeps $\degin_I(r)=0$ and $\degin_I(v)\le k$ for every $v\ne r$, and maintains
\begin{equation}\label{eq:price-invariant}
	I\in\arg\min\{a(X):X\text{ is independent in }\mathcal G^k\}.
\end{equation}
The minimum ranges over all independent sets, with no restriction on their cardinalities or indegrees. Since $a(X)=c(X)-\sum_v h(v)\degin_X(v)$, sets with the same indegree vector receive the same price adjustment. Thus this invariant implies that $I$ is extreme within its indegree vector as well.

We use the following standard local optimality conditions to maintain this invariant.

\begin{lemma}[Local optimality conditions]
	\label{lem:local-optimality}
	Let $M$ be a matroid with costs $a$ on its ground set, and let $I$ be independent. Then $I$ minimizes $a$ over all independent sets iff
	\begin{align}
		a(f) & \le 0    &  & \text{for } f\in I,\label{eq:opt-selected}                                                                                           \\
		a(e) & \ge 0    &  & \text{for } e\notin I \text{ with } I+e\text{ independent},\label{eq:opt-free}                                                       \\
		a(e) & \ge a(f) &  & \text{for } e\notin I \text{ with } I+e\text{ dependent, and } f\in I \text{ with } I-f+e\text{ independent}.\label{eq:opt-exchange}
	\end{align}
\end{lemma}
\begin{proof}
	Necessity is clear. For sufficiency, run the matroid greedy algorithm on the negative-cost elements in increasing cost order, breaking ties in favor of elements of $I$. Inductively, it accepts exactly the elements of $I$ encountered so far: an element of $I$ can be added since $I$ is independent; for a negative-cost $e\notin I$, \eqref{eq:opt-free} implies that $I+e$ is dependent, and \eqref{eq:opt-exchange} puts every element of $I$ in its fundamental circuit before $e$ in the order, so $e$ is rejected. The resulting set is optimal among all independent sets. By~\eqref{eq:opt-selected}, the remaining elements of $I$ have cost zero; adding them preserves optimality and gives $I$.
\end{proof}

By \Cref{lem:local-optimality}, \eqref{eq:price-invariant} holds iff the conditions \eqref{eq:opt-selected}--\eqref{eq:opt-exchange} hold for the adjusted costs $a$.

\paragraph{Initialization.} If some $v\ne r$ has fewer than $k$ entering arcs, no packing exists, and we stop. Otherwise start with $I=\varnothing$, $h(v)=\min\{c(e):\head(e)=v\}$ for $v\ne r$, and $h(r)=0$. Then all adjusted costs are nonnegative, and~\eqref{eq:price-invariant} holds.

\paragraph{Adjusted distances.} For $v\in V$, let $d_v$ be the minimum cost of an augmenting path from $v$ in $D(I)$, with $d_v=+\infty$ if none exists, and define the \emph{adjusted distance} $\delta_v=d_v-h(v)$. For an augmenting path $P$, write $a(P)=a(I\oplus P)-a(I)$ for its adjusted cost. If $P$ starts from $u$, the prices of $f_j$ and $e_{j+1}$ cancel as they share the same head, so
\begin{equation}\label{eq:adjusted-path-cost}
	a(P)=\sum_ja(e_j)-\sum_ja(f_j)=\ell(P)-h(u).
\end{equation}
Regrouping the element costs gives $a(P)=\sum_{j=0}^{q-1}(a(e_j)-a(f_j))+a(e_q)$. By \Cref{lem:local-optimality}, every term is nonnegative, so $\delta_v\ge 0$.

\subsection{Price Update}\label{sec:price-update}

Assume that the augmenting distances $d_v$ and shortest proposals have been computed. A shortest proposal from $u$ has adjusted cost $\delta_u$. If $\delta_u>0$, executing it would violate the price invariant: the old independent set $I$ would have smaller adjusted cost than the new one. 

We therefore first update the prices so that $I$ remains optimal and every shortest proposal has zero adjusted cost. Any compatible batch then preserves the same optimum. We denote the new prices by $h^+$ and the new adjusted costs by $a^+(e)\coloneqq c(e)-h^+(\mathrm{head}(e))$.

By~\eqref{eq:adjusted-path-cost}, setting $h^+(u)=d_u$ makes a shortest proposal from $u$ have adjusted cost $d_u-h^+(u)=0$. To handle vertices with no augmenting path as well, we cap the increments:
\[
	L=\max\bigl(\{0\}\cup\{\delta_v:\delta_v<+\infty\}\bigr),
	\qquad
	h^+(v)=h(v)+\min\{\delta_v,L\}.
\]
Since $L$ is the largest finite $\delta_v$, the cap only changes vertices with $\delta_v=+\infty$: it keeps their prices finite, and it is large enough for the exchange conditions in the proof below.

\begin{lemma}[Price update and batch augmentation]\label{lem:price-update}\label{lem:batch-price-invariant}
	\leavevmode
	\begin{enumerate}[label=(\alph*)]
		\item $I$ minimizes $a^+$ over all independent sets of $\mathcal G^k$.
		\item If $\delta_u<+\infty$, then $h^+(u)=d_u$, and every shortest augmenting path $P$ from $u$ has $a^+(P)=0$.
		\item After executing the proposals whose terminals belong to a compatible set $A'$, the resulting set $I'$ minimizes $a^+$ over all independent sets of $\mathcal G^k$.
	\end{enumerate}
\end{lemma}
\begin{proof}
	(a) Write $\gamma_v=h^+(v)-h(v)=\min\{\delta_v,L\}$ for the price increment at $v$. Then $0\le\gamma_v\le L$ and $a^+(e)=a(e)-\gamma_{\head(e)}$. We verify the three conditions of \Cref{lem:local-optimality}.
	\begin{itemize}
		\item For $f\in I$, increasing the price can only decrease its adjusted cost, so $a^+(f)\le a(f)\le0$.
		\item For a free element $e$, the single-element augmenting path $(e)$ has adjusted cost $a(e)$. Hence $\gamma_{\head(e)}\le\delta_{\head(e)}\le a(e)$, giving $a^+(e)\ge0$.
		\item Consider a nonfree $e\notin I$ and $f\in I$ with $I-f+e$ independent. Given an augmenting path $P$ from $\head(f)$, prepend the $\mathcal{G}^k$ exchange $e\to f$ and a partition exchange from $f$ to the first element of $P$. Deleting nonnegative cycles gives an augmenting path from $\head(e)$. Hence by the triangle inequality of shortest paths,
		      \[
			      \delta_{\head(e)}\le a(e)-a(f)+\delta_{\head(f)}.
		      \]
		      Since $a(e)-a(f)\ge0$, taking the capped distances gives
		      \[
			      \gamma_{\head(e)}
			      \le \min\{a(e)-a(f)+\delta_{\head(f)},L\}
			      \le a(e)-a(f)+\gamma_{\head(f)}.
		      \]
		      Thus $a^+(e)\ge a^+(f)$.
	\end{itemize}
	Thus $I$ remains optimal for $a^+$.

	(b) If $\delta_u$ is finite, then $h^+(u)=h(u)+\delta_u=d_u$. By~\eqref{eq:adjusted-path-cost}, every shortest augmenting path $P$ from $u$ therefore has
	\[
		a^+(P)=\ell(P)-h^+(u)=d_u-d_u=0.
	\]

	(c) By~(a), the old set $I$ is optimal for $a^+$. By~(b), each executed proposal has zero adjusted cost. Since the executed proposals are element-disjoint,
	\[
		a^+(I')=a^+(I)+\sum_{T:e_T\in A'}a^+(P_T)=a^+(I).
	\]
	The new set $I'$ is independent by \Cref{lem:batch-optimality}, so it attains the same optimum.
\end{proof}

Parts~(a) and~(b) also give an alternative price-based proof of the lower bound in \Cref{thm:relaxation}(b).

\subsection{Potential for the Compressed Graph}\label{sec:feasible-potential}

We now extend the vertex prices to a feasible potential on the compressed graph. Let $Q$ be the subgraph of $\CG$ induced by the vertices for elements of $I$ and the jump vertices, with only the arcs that represent cyclic forest swaps. Define
\begin{equation}\label{eq:feasible-potential}
	\begin{aligned}
		p(v^\star) & =h(v),\qquad p(z)=0,                             \\
		p(x)       & =\min\{-a(f): f\in I,\ x\leadsto f\text{ in }Q\}
		\qquad(x\in V(Q)).
	\end{aligned}
\end{equation}
Here reachability includes paths of length zero. The internal potentials propagate the bounds $-a(f)$ backward along scan paths. Write $w_p(x,y)=w(x,y)-p(x)+p(y)$ for the reduced arc weights.

\begin{lemma}[A feasible potential]\label{lem:feasible-potential}
	Under~\eqref{eq:price-invariant}, all reduced weights $w_p$ are nonnegative. In particular, $\CG$ has no negative cycle.
\end{lemma}
\begin{proof}
	Every vertex of $Q$ reaches an element of $I$, so $p$ is finite.
	We verify nonnegativity for each type of arc.
	\begin{itemize}
		\item \emph{Forest-scan arcs $x\to y$.} Every element of $I$ reachable from $y$ is reachable from $x$, so $p(x)\le p(y)$. Hence
		      \[
			      w_p(x,y)=0+p(y)-p(x)\ge0.
		      \]

		\item \emph{Arcs $f\to\head(f)^\star$, $f\in I$.} By the definition of $p$, $p(f)\le -a(f)$ since $f$ reaches itself in $Q$. So
		      \[
			      w_p(f,\head(f)^\star)
			      \ge -c(f)+h(\head(f))+a(f)=0.
		      \]

		\item \emph{Arcs $\head(e)^\star\to x$ labeled by $e$.} Let $f\in I$ achieve the minimum defining $p(x)$, so $p(x)=-a(f)$ and $f$ is reachable from $x$ through forest swaps.

		      If $e$ is nonfree, $I-f+e$ is independent by \Cref{lem:cyclic-reachability}, so $a(e)\ge a(f)$ by~\eqref{eq:opt-exchange}. If $e$ is free, $a(e)\ge0\ge a(f)$ by~\eqref{eq:opt-free} and~\eqref{eq:opt-selected}. Hence
		      \[
			      w_p(\head(e)^\star,x)=c(e)-a(f)-h(\head(e))=a(e)-a(f)\ge0.
		      \]

		\item \emph{Arcs $\head(e)^\star\to z$ labeled by a free element $e$.} Since $p(z)=0$, their reduced weight is
		      \[
			      w_p(\head(e)^\star,z)=c(e)+0-h(\head(e))=a(e)\ge0
		      \]
		      by~\eqref{eq:opt-free}.
	\end{itemize}
\end{proof}

We compute $p$ in $O(M+N)$ time by propagating minima backward over the acyclic graph of strongly connected components of $Q$.

\subsection{Algorithm and Analysis}\label{sec:priced-implementation}\label{sec:main-theorem-proof}

We first show how one Dijkstra computation gives both the adjusted
distances needed for the price update and the proposals needed for
batch augmentation.

\begin{lemma}[Computing Adjusted Distances and Proposals]\label{lem:dijkstra-search}
	Under~\eqref{eq:price-invariant}, the distance from $v^\star$ to $z$ in $\CG$ with weights $w_p$ is $\delta_v$. One reverse Dijkstra computation from $z$, followed by path recovery as in \Cref{sec:compressed-search-graph}, computes all $\delta_v$ and the proposals for the chosen vertices with finite distance in $O(m+kn\log n\log(kn))$ total time.
\end{lemma}
\begin{proof}
	By \Cref{lem:feasible-potential}, all reduced weights are nonnegative.
	For any $v^\star$-to-$z$ path $P$, the potential terms cancel along the path, and 
	\[
		w_p(P)=w(P)-p(v^\star)+p(z)=w(P)-h(v).
	\]
	By \Cref{lem:compressed-graph-search}, minimizing $w(P)$ gives
	the minimum cost of an augmenting path from $v$.
	Subtracting $h(v)$ gives its minimum adjusted cost, namely $\delta_v$.
	Thus Dijkstra on the reversed graph computes the adjusted distances at the head
	vertices.

  Since reweighting shifts all path costs from the same start by the same amount and leaves the exchange counts unchanged, the path-recovery procedure of \Cref{sec:compressed-search-graph} applies directly with weights $w_p$ in place of $w$.

	Graph construction, potentials, and recovery take $O(m+kn\log n)$
	time. Using Fibonacci heaps~\cite{fredmanFibonacciHeapsTheir1987},
	Dijkstra takes $O(M+N\log N)=O(m+kn\log n\log(kn))$ time by
	\eqref{eq:compressed-graph-size}.
\end{proof}

Initialize $I=\varnothing$ and the prices as in
\Cref{sec:price-invariant}, and use \Cref{alg:priced-phase} for each
iteration of \Cref{alg:batching-framework}.
The price update uses the distances of all head vertices, including
those not chosen as augmentation starts; unreachable heads receive
the capped increment of \Cref{sec:price-update}.

\begin{algorithm}[ht]
	\caption{One phase with Dijkstra and selection on the contracted graph}\label{alg:priced-phase}
	\begin{algorithmic}[1]
		\Require $I$ with a forest decomposition and prices satisfying~\eqref{eq:price-invariant}
		\State compute $\cT(I)$; choose a deficient vertex $u_T$ in each deficient block
		\State build $\CG$ with cyclic scans and compute $p$
		\State run Dijkstra from $z$ in the reverse of $\CG$ with weights $w_p$; let $d_p$ be the distances
		\State set $\delta_v\gets d_p(v^\star)$ for all $v\in V$
		\State if some $\delta_{u_T}=+\infty$, return ``infeasible''
		\State recover the proposals $P_T$ using distances $d_p$ and weights $w_p$, as in \Cref{sec:compressed-search-graph}
		\State $L\gets\max(\{0\}\cup\{\delta_v:\delta_v<\infty\})$; $h(v)\gets h(v)+\min\{\delta_v,L\}$ for all $v$
		\State contract the old blocks and form the arc set $H=\Ix\dotcup A$ using the old $I$ and the proposed terminals
		\State $K\gets\textsc{Extend}(H)$; $A'\gets K\setminus \Ix$
		\State $I\gets I\oplus\displaystyle\bigcup_{T:e_T\in A'}P_T$
		\State compute a forest decomposition of $I$ using \Cref{lem:ak-forest-packing}
		\State return $I,h$ with these forests
	\end{algorithmic}
\end{algorithm}

\begin{proof}[Proof of \Cref{thm:main}]
	\emph{Correctness.} By \Cref{lem:dijkstra-search,lem:price-update},
	our implementation computes the required proposals and preserves
	the price invariant. Correctness therefore follows from the batching
	framework in \Cref{sec:batching-framework}, followed by the
	arborescence extraction of \Cref{lem:arborescence-extraction}.

	\emph{Running time.} By \Cref{lem:dijkstra-search} and
	\Cref{sec:compatible-set-selection}, we may take
	$\mathsf S(N,M)=O(m+kn\log n\log(kn))$ and
	$\Sigma=O(k^3n\log(kn))$ in
	\Cref{lem:parametric-running-time}, giving~\eqref{eq:main-time}.
	\Cref{tab:phase-costs} summarizes the costs of each step.

	\begin{table}[H]
		\centering\small
		\begin{tabular}{@{}ll@{}}
			\toprule
			Step                                                           & Time per phase            \\
			\midrule
			Tight blocks and contracted graph                              & $O(kn\log n)$             \\
			Compressed graph and feasible potential                        & $O(m+kn\log n)$           \\
			Dijkstra, tie-breaking, and path recovery                      & $O(m+kn\log n\log(kn))$   \\
			Basis computation and mandatory repair on the contracted graph & $O(k^3t\log(kn))$         \\
			Apply chosen proposals and rebuild on the original vertices    & $O(k^3n\log(kn))$         \\
			\midrule
			Number of phases                                               & $O(k\log(kn))$            \\
			Final arborescence extraction (once)                           & $O(km\log n+k^4n\log^2n)$ \\
			\bottomrule
		\end{tabular}
		\caption{Cost of each step, where $t\le n$ is the number of old tight blocks. Summed over the phases, the $N\log N$ term of Dijkstra's algorithm gives the $k^2n\log n\log^2(kn)$ term of~\eqref{eq:main-time}.}
		\label{tab:phase-costs}
	\end{table}

\end{proof}

\section{Applications}\label{sec:applications}

\subsection{Indegree-Bounded Forest Packing}\label{sec:indegree-bounded-forest-packing}

\begin{theorem}[Indegree-bounded forest packing]\label{thm:indegree-bounded-forest-packing}
	Let $\bar b_v\in\{0,\ldots,k\}$ for every $v\in V$. Call $I\subseteq E$ \emph{feasible} if $I$ is independent in $\mathcal G^k$ and $\degin_I(v)\le\bar b_v$ for every $v$. A feasible set of maximum cardinality and, subject to that, of minimum cost can be computed deterministically in
	\[
		O(km\log(kn)+k^2n\log n\log^2(kn)+k^4n\log^2(kn))
	\]
	time for arbitrary real costs. The algorithm also returns a forest decomposition of the set.
\end{theorem}

\begin{proof}
	The algorithm in \Cref{thm:main} requires indegree exactly $k$ at every nonroot vertex. We add a new root $s$ with $k$ arcs into each vertex $v$, of two kinds:

	\begin{itemize}
		\item $k-\bar b_v$ \emph{forced} arcs $s\to v$ of cost $-(kn+1)M$. They are so cheap that every optimum contains all of them, so they use up the capacity that the bound $\bar b_v$ takes away.
		\item $\bar b_v$ \emph{heavy} arcs $s\to v$ of cost $M$. They serve as placeholders for unused slots; they are so expensive that an optimum uses as few of them as possible.
	\end{itemize}
	Here $M=1+\sum_{e\in E}|c(e)|$ exceeds the total absolute cost of the original arcs, and we denote the set of $kn$ new arcs by $E_s$. Then $E_s$ alone already forms a packing of $k$ $s$-arborescences, so we can apply \Cref{thm:main} to the auxiliary graph $G'=(V\cup\{s\},E\cup E_s)$ to get a minimum-cost packing $I'$.

	On the one hand, $I'$ gives a feasible set in $G$.
	First, $I'$ contains all forced arcs: omitting one increases their total cost by at least $(kn+1)M$, while the total cost of all other arcs can vary by at most $knM+\sum_e|c(e)|<(kn+1)M$. Thus any packing omitting a forced arc is more expensive than the packing consisting of all arcs in $E_s$. Let $I=I'\cap E$. It is independent as a subset of $I'$, and the forced arcs leave only $\bar b_v$ slots at each $v$, so $\degin_I(v)\le\bar b_v$. Hence $I$ is feasible in $G$. The packing $I'$ uses $\sum_v\bar b_v-|I|$ heavy arcs, so
	\[
		c(I')=\text{constant}+c(I)-M|I|,
	\]
	where the constant depends only on the instance.

	On the other hand, any feasible set $J\subseteq E$ can be extended to a packing in $G'$. Add all forced arcs and $\bar b_v-\degin_J(v)$ heavy arcs at each $v$, obtaining a set $J'$ with indegree $k$ at every original vertex. To check~\eqref{eq:forest-density}, a vertex set not containing $s$ contains only arcs of $J$, while a set $S$ containing $s$ has at most $k(|S|-1)$ internal arcs by the indegree bound. Hence $J'$ is independent in $\mathcal G^k$, and \Cref{lem:arborescence-formulation} shows that it is a packing of $k$ $s$-arborescences. This extension uses $\sum_v\bar b_v-|J|$ heavy arcs, so
	\[
		c(J')=\text{constant}+c(J)-M|J|,
	\]
	with the same constant as above.

	Finally, by the optimality of $I'$, we have $c(I')\le c(J')$, so
	\[
		c(I)-M|I|\le c(J)-M|J|
	\]
	for every feasible $J$. The costs of any two subsets of $E$ differ by less than $M$, so $I$ has maximum cardinality and, subject to that, minimum cost.

	The auxiliary graph has $n+1$ vertices and $m+kn$ arcs, so \Cref{thm:main} gives the stated running time.
\end{proof}

\subsection{Undirected Forest Packing}\label{sec:undirected-forest-packing}

In this subsection $G=(V,E)$ is an undirected multigraph. The matroid $\mathcal G^k$ on edge set $E$ is the union of $k$ copies of the graphic matroid. Its bases are the maximum-cardinality unions of $k$ forests, so the task is to find a minimum-cost basis of $\mathcal G^k$.

Write $\mathsf{MCF}(N,M)$ for the time of a deterministic algorithm that finds an integral minimum-cost maximum flow in a network with $N$ vertices and $M$ arcs, whose costs and capacities are integers bounded by a polynomial in $N+M$.

\undirectedforestpacking*

Plugging in deterministic min-cost flow~\cite{vandenbrandDeterministicAlmostLinear2023} gives the running time
	\begin{equation}\label{eq:undirected-forest-time}
		O\bigl(k\log(kn)\,(m+n)^{1+o(1)}
		+k^3n\log n\log^3(kn)+k^5n\log^3(kn)\bigr).
	\end{equation}
	The resulting algorithm is strongly polynomial and runs in $(m+n)^{1+o(1)}$ time for fixed $k$.

Before describing the algorithm, we introduce pseudoforests.

\paragraph{Pseudoforests.} A \emph{pseudoforest} is a graph in which every connected component has at most one cycle. Let $\mathcal P$ be the pseudoforest matroid, also called the bicircular matroid, whose independent sets are the pseudoforests, and let $\mathcal P^k$ be the union of $k$ copies of $\mathcal P$; its independent sets are the unions of $k$ edge-disjoint pseudoforests, which we call \emph{$k$-pseudoforests}. A set $X$ is independent in $\mathcal P^k$ iff
\begin{equation}\label{eq:pseudoforest-density}
	|X[U]|\le k|U|\qquad\text{for every }U\subseteq V.
\end{equation}
Equivalently, $X$ can be oriented so that every vertex has indegree at most $k$.

Indeed, by Hakimi's orientation theorem~\cite[Theorem~4]{HAKIMI1965290}, \eqref{eq:pseudoforest-density} holds iff $X$ has an orientation with indegree at most $k$ at every vertex. Given such an orientation, split $X$ into $k$ parts so that every vertex has at most one incoming edge in each part. Each part has indegree at most one, so it is a pseudoforest. Conversely, orienting each pseudoforest with indegree at most one gives an orientation of $X$ with indegree at most $k$. Since $k(|U|-1)\le k|U|$, every $\mathcal G^k$-independent set is $\mathcal P^k$-independent.

\paragraph{Overview.} The algorithm repeats a round that follows~\cite[Section~4]{arkhipovFasterAlgorithmsPacking2025}.
\begin{enumerate}[label=(\arabic*)]
	\item Choose a minimum-cost maximum-cardinality $k$-pseudoforest $\Pi$, together with an orientation of indegree at most $k$. This can be done with minimum-cost maximum flow.
	\item Inside $\Pi$, choose a minimum-cost maximum-cardinality union $\Phi$ of $k$ forests. Using this orientation, we find $\Phi$ by \Cref{thm:indegree-bounded-forest-packing} with all bounds equal to $k$.
	\item Contract all $\Phi$-tight blocks, and repeat on the contracted graph. The edges of $\Phi$ that survive the contraction are \emph{inherited} by the next round.
\end{enumerate}
Contracting a block fixes its internal forest edges in the final solution, so we must ensure that these edges belong to an optimal solution. The next lemma shows that if both choices follow the same greedy order, then $\Phi$ is contained in the forest packing obtained by running greedy directly on all edges under that order.

\subsubsection{Two-Stage Greedy Lemma}\label{sec:two-stage-greedy}

\paragraph{Greedy bases.} Fix a strict total order on $E$. For a matroid $M$ on $E$ and a set $X\subseteq E$, let $\operatorname{Greedy}_M(X)$ be the set obtained by scanning $X$ in this order and accepting every edge that keeps the accepted set independent. We use three standard facts.
\begin{itemize}
	\item If $Y$ is a prefix of the order, then $\operatorname{Greedy}_M(X)\cap Y=\operatorname{Greedy}_M(X\cap Y)$.
	\item An edge is rejected iff it lies in the closure of the edges accepted before it. In particular, $\operatorname{Greedy}_M(X)$ is a basis of $X$ in $M$.
	\item If the order lists the edges by nondecreasing cost, then $\operatorname{Greedy}_M(X)$ is a minimum-cost basis of $X$. It is the unique one if all costs are distinct.
\end{itemize}
Order the edges of $G$ by $(c(e),\text{identifier}(e))$, and call this the \emph{cost order}. Let $B^\ast=\operatorname{Greedy}_{\mathcal G^k}(E)$ for the cost order. By the third fact, $B^\ast$ is a minimum-cost maximum-cardinality union of $k$ forests, and it is the set that \Cref{thm:undirected-forest-packing} computes. We call $B^\ast$ the \emph{greedy optimum}.

\begin{lemma}[Two-stage greedy]\label{lem:two-stage-greedy}
	Fix a strict total order on $E$. With respect to this order, let
	\[
		\Pi=\operatorname{Greedy}_{\mathcal P^k}(E),\qquad
		\Phi=\operatorname{Greedy}_{\mathcal G^k}(\Pi),\qquad
		B=\operatorname{Greedy}_{\mathcal G^k}(E).
	\]
	Then $\Phi\subseteq B$.
\end{lemma}
\begin{proof}
	Run the three greedy scans in parallel. For a prefix $X$ of the order, write $B_X=B\cap X$, $\Pi_X=\Pi\cap X$, and $\Phi_X=\Phi\cap X$. By the first two facts, $B_X$ and $\Pi_X$ are bases of $X$ in $\mathcal G^k$ and in $\mathcal P^k$, and $\Phi_X$ is a basis of $\Pi_X$ in $\mathcal G^k$. We show that for every prefix $X$,
	\begin{equation}\label{eq:two-stage-span}
		\cl_{\mathcal G^k}(B_X)
		\subseteq \cl_{\mathcal P^k}(\Pi_X)\cup\cl_{\mathcal G^k}(\Phi_X).
	\end{equation}
	Intuitively, the left-hand side denotes all the elements that will be ``rejected'' by $B_X$, while the right-hand side denotes that these elements should be either ``rejected'' by $\Pi_X$ (so they won't have the chance of getting into $\Phi$), or ``rejected'' by $\Phi_{X}$ itself.

	\emph{The closure in $\mathcal P^k$.} Call a vertex set $U$ \emph{saturated} if $|\Pi_X[U]|=k|U|$, that is, if $U$ attains the bound~\eqref{eq:pseudoforest-density}. The function $U\mapsto|\Pi_X[U]|$ is supermodular, so~\eqref{eq:pseudoforest-density} for $U\cap W$ gives $|\Pi_X[U\cup W]|\ge k|U|+k|W|-k|U\cap W|=k|U\cup W|$ for saturated $U$ and $W$. Hence unions of saturated sets are saturated, and the union $Z$ of all saturated sets is the largest saturated set; $Z$ may be empty. An edge $e\notin\Pi_X$ makes $\Pi_X+e$ dependent iff some saturated set contains both ends of $e$, that is, iff $e\in E[Z]$. Therefore
	\[
		\cl_{\mathcal P^k}(\Pi_X)=\Pi_X\cup E[Z].
	\]

	\emph{Blocks of $B_X$ do not cross $Z$.} Let $T$ be a $B_X$-tight block, so $|B_X[T]|=k(|T|-1)$. We claim that $T\subseteq Z$ or $T\cap Z=\varnothing$. Suppose not. Since $B_X\subseteq X\subseteq\cl_{\mathcal P^k}(\Pi_X)$, every edge of $B_X[T]$ outside $E[Z]$ belongs to $\Pi_X$. Using~\eqref{eq:forest-density} for $B_X$ and the nonempty set $T\cap Z$, we get
	\begin{align*}
		|\Pi_X[Z\cup T]|
		 & \ge |\Pi_X[Z]|+(|B_X[T]|-|B_X[T\cap Z]|) \\
		 & \ge k|Z|+(k(|T|-1)-k(|T\cap Z|-1))
		=k|Z\cup T|.
	\end{align*}
	By~\eqref{eq:pseudoforest-density}, equality holds, so $Z\cup T$ is saturated. Since $T\not\subseteq Z$, this contradicts the maximality of $Z$.

	\emph{Proof of~\eqref{eq:two-stage-span}.} Let $e\in\cl_{\mathcal G^k}(B_X)\setminus\cl_{\mathcal P^k}(\Pi_X)$. This is an edge that the direct scan would reject but the first stage would accept. We show that $e\in\cl_{\mathcal G^k}(\Phi_X)$, so the second stage would reject it. Since $X\subseteq\cl_{\mathcal P^k}(\Pi_X)$, we have $e\notin X$, and in particular $e\notin B_X$. As $B_X+e$ is dependent, \Cref{lem:free-elements} puts both ends of $e$ in one $B_X$-tight block $T$, and $e\in\cl_{\mathcal G^k}(B_X[T])$. The block $T$ is not inside $Z$, since otherwise $e\in E[Z]\subseteq\cl_{\mathcal P^k}(\Pi_X)$. By the claim, $T$ is disjoint from $Z$, so $B_X[T]\subseteq\Pi_X$. Since $\Phi_X$ is a basis of $\Pi_X$ in $\mathcal G^k$,
	\[
		e\in\cl_{\mathcal G^k}(B_X[T])\subseteq\cl_{\mathcal G^k}(\Pi_X)=\cl_{\mathcal G^k}(\Phi_X).
	\]

	\emph{Conclusion.} Let $e\in\Phi$, and let $X$ be the set of edges before $e$. Both stages accept $e$, so $e\notin\cl_{\mathcal P^k}(\Pi_X)$ and $e\notin\cl_{\mathcal G^k}(\Phi_X)$. By~\eqref{eq:two-stage-span}, $e\notin\cl_{\mathcal G^k}(B_X)$, so the direct scan accepts $e$. Thus $\Phi\subseteq B$.
\end{proof}

\begin{remark}[Matroid pairs]
	The conclusion of \Cref{lem:two-stage-greedy} extends to any pair of matroids $M,N$ on a common ground set $E$, with $M,N$ in place of $\mathcal G^k,\mathcal P^k$, respectively, provided that
	\[
		\cl_M(\cl_N(S))=\cl_N(S)\cup\cl_M(S)
		\qquad\text{for every $N$-independent set $S$.}
	\]
\end{remark}

\subsubsection{Algorithm and Analysis}\label{sec:undirected-algorithm}

\paragraph{Preprocessing.} For each pair of vertices, keep only the $k$ cheapest parallel edges under the cost order. As every union of $k$ forests contains at most $k$ edges between any two vertices, this does not change the solution. The number of edges is thus reduced to at most $kn^2$, which is polynomial in $n$ and $k$; this absorbs the cost sorting time and keeps the running-time bound clean.

Let $R$ be the set inherited from the preceding round, initially empty. Let $\prec_R$ list $R$ first in the cost order, followed by the remaining edges in the cost order. In each round we compute
\[
	\Pi=\operatorname{Greedy}_{\mathcal P^k}(E),\qquad
	\Phi=\operatorname{Greedy}_{\mathcal G^k}(\Pi)
\]
with respect to $\prec_R$, contract the $\Phi$-tight blocks, and inherit
the remaining edges of $\Phi$.

\begin{algorithm}[htbp]
	\caption{Minimum-cost maximum-cardinality undirected forest packing}\label{alg:undirected-forest-packing}
	\begin{algorithmic}[1]
		\Require undirected multigraph $G$, integer $k\ge1$, real edge costs $c$
		\Ensure a minimum-cost maximum-cardinality union of $k$ forests, with a forest decomposition
		\State delete loops; for each vertex pair, keep only its $k$ cheapest parallel edges
		\State sort the remaining edges by $(c(e),\text{identifier}(e))$; $R\gets\varnothing$; start with an empty stack
		\Loop
		\State compute $\Pi=\operatorname{Greedy}_{\mathcal P^k}(E)$ with respect to $\prec_R$, with an orientation of indegree $\le k$

		\State compute $\Phi=\operatorname{Greedy}_{\mathcal G^k}(\Pi)$ with respect to $\prec_R$, with a forest decomposition
		\If{$|\Phi|=|R|$}
		\State lift the forest decomposition of $\Phi$ through previous contractions, and \Return it
		\EndIf
		\State compute $\cT(\Phi)$ and let $K$ be the set of edges of $\Phi$ inside blocks; push the $k$ trees inside every nonsingleton block onto the stack
		\State contract every block, delete loops, and set $R\gets\Phi\setminus K$
		\EndLoop
	\end{algorithmic}
\end{algorithm}

We implement the two greedy choices using minimum-cost algorithms. As in~\cite[Section~3]{gabowForestsFramesGames1992}, assign each edge $e$ the cost $\operatorname{pos}(e)$, its position in $\prec_R$. Since these costs are distinct and increase along $\prec_R$, the unique minimum-cost bases are exactly the bases obtained by greedy in this order.

\paragraph{Computing $\Pi$.} We build a flow network with a super source $\mathsf{src}$, a super sink $\mathsf{snk}$, a node $x_e$ for every current edge $e=\{u,v\}$, and a node for every current vertex. Its arcs are as follows.
\begin{center}
	\begin{tabular}{@{}lcc@{}}
		\toprule
		arc                   & capacity            & cost                    \\
		\midrule
		$\mathsf{src}\to x_e$ & $1$                 & $\operatorname{pos}(e)$ \\
		$x_e\to u,\ x_e\to v$ & $1$                 & $0$                     \\
		$v\to\mathsf{snk}$    & $\min\{k,\deg(v)\}$ & $0$                     \\
		\bottomrule
	\end{tabular}
\end{center}
An integral flow selects $e$ when $\mathsf{src}\to x_e$ carries a unit, and it orients $e$ toward the end that receives this unit; this gives a $k$-pseudoforest with an orientation of indegree at most $k$. The edge set selected by a maximum flow is therefore a basis of $\mathcal P^k$, and a minimum-cost maximum flow gives the minimum-$\operatorname{pos}$-cost basis, which is $\Pi$, together with an orientation. The network has $O(m+n)$ vertices and arcs, and its costs and capacities are integers at most $m$.

\paragraph{Computing $\Phi$.} Apply \Cref{thm:indegree-bounded-forest-packing} to this orientation of $\Pi$, with costs $\operatorname{pos}$ and indegree bounds
$\bar b_v=k$ for every vertex $v$. Since the orientation has indegree at most $k$, every subset of $\Pi$ satisfies the bounds, so the indegree bounds impose no extra restriction here. Hence we obtain $\Phi$ as the minimum-$\operatorname{pos}$-cost maximum-cardinality $k$-forest inside $\Pi$, together with a forest decomposition.

\begin{proof}[Proof of \Cref{thm:undirected-forest-packing}]
	\emph{Optimality.} At the start of each round, let $B^\ast$ be
	the greedy optimum of the current graph under the cost order.
	We maintain two invariants:
	\begin{enumerate}
		\item $R\subseteq B^\ast$;
		\item adding the edges saved inside previously contracted
		      blocks to $B^\ast$ gives the greedy optimum of the original graph.
	\end{enumerate}
	Both hold initially, since $R=\varnothing$ and no edges have
	been saved.

	We use a simple property of greedy: moving a subset of a greedy basis to the front, while preserving the order of all other edges, leaves that basis unchanged. Indeed, inductively, every basis edge is still accepted, and every other edge is still rejected because the earlier basis edges that caused its rejection remain before it.
	Since $R\subseteq B^\ast$, greedy under $\prec_R$ still returns $B^\ast$. Both scans accept $R$ first, since it is independent in both matroids. Together with \Cref{lem:two-stage-greedy}, this gives
	\begin{equation}\label{eq:inherited-edges-invariant}
		R\subseteq\Phi\subseteq B^\ast.
	\end{equation}

	Now consider the contraction step. Let $K=\bigcup_{T\in\cT(\Phi)}\Phi[T]$ be the edges contracted in this round. Since $K\subseteq\Phi\subseteq B^\ast$, all these edges belong to the optimum we want to recover. We claim that the greedy optimum after contraction is $B^\ast\setminus K$. 

  To see this, consider a greedy scan that processes $K$ first and then the remaining edges in the cost order. By the same property, it still returns $B^\ast$. After accepting $K$, it rejects every other edge internal to a block, since $\Phi[T]$ already consists of $k$ spanning trees of $T$. On the surviving edges, its independence tests are exactly those in the contracted graph, by the contraction and lifting argument of \Cref{lem:contraction-lifting}. Hence the claim follows.

	Therefore, the inherited set in the next round is $\Phi\setminus K$, which is contained in $B^\ast\setminus K$, preserving the first invariant. Saving the edges of $K$ and replacing $B^\ast$ by $B^\ast\setminus K$ also preserves the second.

	\emph{Termination.} Ignoring costs, each round follows the same steps as
	\cite[Algorithm~4]{arkhipovFasterAlgorithmsPacking2025}:
	$\Pi$ is a maximum-cardinality $k$-pseudoforest containing $R$,
	$\Phi$ is a maximum-cardinality forest packing inside $\Pi$,
	and all $\Phi$-tight blocks are contracted.
	The cost-based choices preserve these properties, so the
	progress analysis of~\cite[Lemmas~4.5--4.6 and Corollary~4.7]
	{arkhipovFasterAlgorithmsPacking2025} applies:
	there are $O(k\log(kn))$ rounds, and when $|\Phi|=|R|$,
	the set $\Phi$ has maximum cardinality in the current graph.

	Since $\Phi\subseteq B^\ast$, we have $\Phi=B^\ast$ at termination.
	By the second invariant, lifting its forest decomposition through
	the saved trees, as in \Cref{lem:contraction-lifting}, recovers the
	original greedy optimum.

	\emph{Running time.} Preprocessing takes $O(m+n)$ time. The remaining graph has at most $k\binom n2$ edges, so sorting costs $O(m\log(kn))$. Since $\mathsf{MCF}(N,M)=\Omega(N+M)$, these costs are absorbed by the flow term in~\eqref{eq:undirected-modular-time}. In each round we compute $\Pi$ in $\mathsf{MCF}(O(m+n),O(m+n))$ time and, since $|\Pi|\le kn$, compute $\Phi$ in $O(k^2n\log n\log^2(kn)+k^4n\log^2(kn))$ time by \Cref{thm:indegree-bounded-forest-packing}. The remaining work per round takes $O(m+n+kn\log n)$ time, including tight-block computation by \Cref{lem:tight-block-computation}. Since $\sum_T(|T|-1)\le n-1$ over all contracted nonsingleton blocks, storing and lifting the saved trees takes $O(kn)$ time in total. Summing over $O(k\log(kn))$ rounds gives \eqref{eq:undirected-modular-time}.

	All flow costs and capacities are integers bounded by $m$, so using~\cite{vandenbrandDeterministicAlmostLinear2023} gives \eqref{eq:undirected-forest-time}. The original real costs are used only in preprocessing and sorting, so the algorithm is strongly polynomial.
\end{proof}

\subsection{Minimum-Cost \texorpdfstring{$k$}{k}-Edge-Connected Spanning Subgraphs}\label{sec:network-design}

For the undirected problem, we use the reduction of Khuller and Vishkin~\cite{khullerBiconnectivityApproximationsGraph1994} (see also~\cite[Theorem~34]{chalermsookApproximatingKEdgeConnected2022}). For the directed problem, we combine an outward and an inward packing. Write $\delta_H(S)$ for the edges of an undirected graph $H$ with exactly one endpoint in $S$, and $\delta^-_H(S)$ for the arcs of a digraph $H$ entering $S$ from outside. An undirected graph $H$ is \emph{$k$-edge-connected} if $|\delta_H(S)|\ge k$ for every nonempty proper vertex set $S$, and a digraph $H$ is \emph{strongly $k$-edge-connected} if $|\delta^-_H(S)|\ge k$ for every such $S$. Costs are nonnegative.

\begin{corollary}\label{cor:network-design}
	Let $n\ge2$. There are deterministic $2$-approximation algorithms, which also detect infeasibility, for
	\begin{enumerate}[label=(\roman*)]
		\item the minimum-cost $k$-edge-connected spanning subgraph of an undirected graph, and
		\item the minimum-cost strongly $k$-edge-connected spanning subgraph of a digraph.
	\end{enumerate}
	Both algorithms have the same running time as the algorithm in \Cref{thm:main}.
\end{corollary}
\begin{proof}
	Each reduction makes at most two packing calls on $n$ vertices and at most $2m$ arcs.

	(i) Fix a root $r$, replace every edge $e=\{u,v\}$ by two arcs $(u,v)$ and $(v,u)$ of cost $c(e)$, compute an optimal packing $P$ of $k$ $r$-arborescences, and return the edges with at least one arc in $P$. For a cut $S$ with $r\notin S$ (possibly after complementing), each arborescence has an arc entering $S$, so $P$ has at least $k$ such arcs. They correspond to distinct undirected edges, because the two orientations of an edge cannot both enter $S$. 

  For an optimal solution $H^*$, the bidirected graph of $H^*$ contains $k$ arc-disjoint $r$-arborescences by Edmonds' theorem~\cite{edmondsEdgeDisjointBranchings1973}, so $c(P)\le2c(H^*)$. The same argument shows that infeasibility of the packing implies infeasibility of the problem.

	(ii) Return the union of an optimal packing $P^+$ of $k$ out-arborescences rooted at $r$ and an optimal packing $P^-$ of $k$ in-arborescences rooted at $r$, the latter computed on the reversed graph. If $r\notin S$, then $P^+$ gives $k$ arcs entering $S$; if $r\in S$, then $P^-$ gives $k$ arcs leaving $V\setminus S$. An optimal $H^*$ contains both kinds of packings by Edmonds' theorem, so $c(P^+\cup P^-)\le c(P^+)+c(P^-)\le2c(H^*)$.
\end{proof}

The same approximation ratio and running time hold for connectivity augmentation: give the edges of the existing subgraph cost zero and add that subgraph to the output.

\appendix

\section{Forest-Swap Reachability and Execution}\label{app:forest-swap-reachability}

We prove \Cref{lem:forest-swap-reachability} for the fixed forest decomposition $I=F_1\dotcup\cdots\dotcup F_k$. We first state triangular exchange for forests and include the elementary proof, following~\cite[Lemma~3.2]{saranurakIncrementalDirectedMinimum2026}.

\begin{lemma}[Triangular exchange for forests]
	\label{lem:triangular-exchange}
	Let $F$ be a forest. For an edge $e$ whose ends are connected in $F$, let $P_F(e)$ denote the path between them. Let $(e_1,f_1),\dots,(e_t,f_t)$ be pairs such that $e_i\notin F$, $f_i\in F$, and $f_i\in P_F(e_i)$ for every $i$. Assume moreover that for every $i<j$, $f_j\notin P_F(e_i)$.
	Then $F+\{e_1,\dots,e_t\}-\{f_1,\dots,f_t\}$ is again a forest.
\end{lemma}
\begin{proof}
	For each $j$, removing $f_j$ from $F$ separates its connected component into two parts. Since $f_j\in P_F(e_j)$, the edge $e_j$ crosses this cut.

	Suppose the final graph contains a cycle $C$, and choose the largest index $j$ such that $e_j\in C$. Since $e_j$ crosses the cut of $F-f_j$, the cycle $C$ must contain another edge crossing the same cut. No edge of $F-f_j$ crosses it. If this other edge is some $e_i$ with $i<j$, then $f_j\in P_F(e_i)$, contradicting the assumption. If it is some $e_i$ with $i>j$, this contradicts the choice of $j$. Hence no such cycle exists, and the final graph is a forest.
\end{proof}

\begin{proof}[Proof of \Cref{lem:forest-swap-reachability}]
    If $e$ is nonfree, its ends lie in one tight block $T$ by \Cref{lem:free-elements}. Every $F_i[T]$ is a spanning tree of $T$, so forest swaps from elements internal to $T$ stay internal to $T$, and no such element is joining. Conversely, suppose that $e$ reaches no joining element, and let $R$ consist of $e$ and the elements of $I$ it reaches. For each $i$, every $g\in R\setminus F_i$ has its $F_i$-path inside $R$, so $F_i\cap R$ spans $R$ in the graphic matroid, and $|F_i\cap R|=\rk_{\mathcal G}(R)$. Hence $|R|=1+k\rk_{\mathcal G}(R)$, so $R$ is not a union of $k$ forests, and $I+e$ is dependent.

    For the second claim, let $e$ be nonfree and let $R$ be as above. If $f$ is not reachable, then $R\subseteq I-f+e$, so $I-f+e$ is dependent. Suppose now that $f$ is reachable. Take a path $e=e_0\to e_1\to\cdots\to e_s=f$ with the fewest forest swaps. For each $j\ge1$, let $F_{i_j}$ be the forest containing $e_j$. We simultaneously replace $e_j$ by $e_{j-1}$ in $F_{i_j}$ for all $j=1,\ldots,s$.

    Fix a forest $F_i$ and list its exchanges as $(x_1,y_1),(x_2,y_2),\ldots$ in path order. Each $F_i-y_a+x_a$ is a forest. For $a<b$, the $F_i$-path of $x_a$ cannot contain $y_b$, since the swap $x_a\to y_b$ would shorten the path. The $x_a$ are distinct and outside $F_i$, and the $y_a$ are distinct and in $F_i$. By \Cref{lem:triangular-exchange}, the resulting edge set is a forest. Each exchange adds and removes an edge within the same original component of $F_i$, so each such component retains its number of edges and remains a spanning tree. Thus each forest keeps its connected components. Intermediate elements move between forests, while $e$ enters and $f$ leaves, so the resulting forests partition $I-f+e$. This proves both the execution claim and the independence of $I-f+e$, completing the characterization of the fundamental circuit.
\end{proof}

\section{Cyclic Scans}\label{app:cyclic-scans}

We prove that the cyclic scans used at the end of \Cref{sec:compressed-search-graph}, a special case of the cyclic-scanning principle of Gabow and Westermann~\cite[Lemma~2.2]{gabowForestsFramesGames1992}, preserve the elements of $I$ reachable from each nonfree element $e\notin I$. This justifies the restriction to one scanned tree path per element.

\begin{lemma}[Cyclic reachability]\label{lem:cyclic-reachability}
	Let $B=F_1\dotcup\cdots\dotcup F_k$ be a union of $k$ spanning trees of a vertex set $T$, and let $e\notin B$ have both ends in $T$. Keep only the following forest swaps:
	\begin{enumerate}[label=(\alph*)]
		\item $e$ scans its path in $F_1$, and
		\item if $k\ge2$, each element of $F_i$ scans its path in $F_{i+1}$ (indices modulo $k$).
	\end{enumerate}
	Then the elements of $B$ reachable from $e$ are exactly the elements of $B$ in the fundamental circuit of $e$ in $\mathcal G^k$.
\end{lemma}
\begin{proof}
	For $k=1$ this is the tree-path description of a fundamental circuit. Let $k\ge2$, and let $R_i\subseteq F_i$ be the set of elements of $F_i$ reachable from $e$. Every element of $R_i$ has its $F_{i+1}$-path in $R_{i+1}$, so
	\[
		\cl_{\mathcal G}(R_1)\subseteq\cl_{\mathcal G}(R_2)\subseteq\cdots\subseteq\cl_{\mathcal G}(R_k)\subseteq\cl_{\mathcal G}(R_1).
	\]
	Thus these closures are equal, with common graphic rank $\rho$. Since each $R_i$ is a forest, it is a basis and $|R_i|=\rho$.

	The $F_1$-path of $e$ lies in $R_1$, so $e\in\cl_{\mathcal G}(R_1)$. Hence $R=\{e\}\cup R_1\cup\cdots\cup R_k$ has size $k\rho+1$ and graphic rank $\rho$. Its rank in $\mathcal G^k$ is therefore at most $k\rho$, so $R$ is dependent.

	On the other hand, $R_1\cup\cdots\cup R_k\subseteq B$ is independent in $\mathcal G^k$. So $R$ contains the unique circuit of $B+e$, the fundamental circuit of $e$, and every element of $B$ in this circuit is reachable from $e$. Conversely, an element reachable with the kept swaps is also reachable with all forest swaps, so it lies in the fundamental circuit by \Cref{lem:forest-swap-reachability}(b) applied to $B$; here $e$ is nonfree since $T$ is $B$-tight.
\end{proof}

We now apply \Cref{lem:cyclic-reachability} to the whole graph $\CG$, in which every element of $E\setminus I$ scans $F_1$ and every element of $F_i$ scans $F_{i+1}$. Fix a nonfree $e\notin I$, and let $T$ be the tight block containing its ends (\Cref{lem:free-elements}). Each $F_i[T]$ is a spanning tree of $T$, so the scanned path of every element internal to $T$ stays in $T$. Moreover, forest swaps only enter elements of $I$, so a path from $e$ never passes through another element of $E\setminus I$. Hence the elements of $I$ reachable from $e$ are those given by \Cref{lem:cyclic-reachability} with $B=I[T]$, namely the elements of $I$ in the fundamental circuit of $e$. This is the reachability property used in \Cref{lem:compressed-graph-search}. Free elements need no such property, since a shortest augmenting path uses a free element only as its last element (\Cref{lem:no-nonterminal-free-elements}), through its arc to $z$. Restricting the scans only removes arcs, so it changes no arc weights or exchange counts.

\section*{Acknowledgments}

\addcontentsline{toc}{section}{Acknowledgments}

The author thanks Thatchaphol Saranurak for suggesting the problem, for recommending textbooks that helped the author develop the virtual-element proof of the relaxation theorem (\cref{thm:relaxation}), and for pointing out that the algorithm of Arkhipov and Kolmogorov could implement the rebuild step. The author also thanks him for many helpful discussions and suggestions on the exposition.

\section*{AI Methodology Statement}

\addcontentsline{toc}{section}{AI Methodology Statement}
The author began studying this problem in fall 2025, before using AI tools for this project. The author developed the initial algorithmic framework, including the idea of extending Gabow's algorithm using matroid-level augmenting paths, the elementary virtual-element argument for the relaxation theorem, much of the batching framework, rebuilding forest decompositions between phases, and sparse-table compression of the search graph.

But the initial batching argument was incomplete. The author had considered tight blocks among possible ways to repair it, but had not resolved how to compute a maximum compatible set of proposals while keeping the required arcs. 

After the release of GPT-6 Astra, the author provided it with a research note describing the above ideas and approach. In subsequent discussions, GPT proposed the approach that developed into the repair procedure in \Cref{sec:compatible-set-selection}. The author further simplified the compatibility argument by observing that tight blocks could be contracted before performing the block-internal replacements.

Following the author's suggestion to replace black-box negative-weight shortest paths with a potential-based method, GPT also proposed the vertex-price update and compressed-graph potential developed in \Cref{sec:price-update,sec:feasible-potential}. The weight assignment and potential were subsequently simplified through further discussion, including the author's suggestion to concentrate the weights on arcs incident to head vertices.

The author devised the reduction for the bounded-indegree problem in \Cref{sec:indegree-bounded-forest-packing}. The author then suggested using ideas from Arkhipov and Kolmogorov to address undirected forest packing; GPT assisted in developing the extension in \Cref{sec:undirected-forest-packing}, particularly the proof of \cref{lem:two-stage-greedy}.

GPT and Claude Opus 5.5 assisted with initial drafting, proof simplifications, and related-work discussions. The author directed the manuscript's organization and exposition, subsequently rewrote the vast majority of the paper, and revised the entire text. The author verified the correctness and originality of all content, including references.

\bibliographystyle{alpha}
\bibliography{refs}

\end{document}